\documentclass[aps,showpacs,preprintnumbers,nofootinbib,amsmath,amssymb]{revtex4-1}
\usepackage{amssymb}
\usepackage{}
\usepackage{slashed,color,amsmath,amssymb,mathrsfs}
\usepackage{graphicx}
\usepackage{ulem}
\usepackage{hyperref}
\usepackage{longtable}
\usepackage{array}

\allowdisplaybreaks

\begin{document}
\preprint{}
\title{Derivation of the gap and Bethe-Salpeter equations at large $N_c$ limit and symmetry preserving truncations}
\author{Hui-Feng Fu}
\email{huifengfu@jlu.edu.cn}
\affiliation{Center for Theoretical Physics, Jilin University,
Changchun 130012, P. R. China}
\author{Qing Wang}
\email[Corresponding author.]{wangq@mail.tsinghua.edu.cn}
\affiliation{Department of Physics, Tsinghua University, Beijing 100084, P. R. China\\
Center for High Energy Physics, Tsinghua University, Beijing 100084, P. R. China\\
Collaborative Innovation Center of Quantum Matter, Beijing 100084, P. R. China}
\author{Libo Jiang}
\email{jiangl@fnal.gov}
\affiliation{Department of Physics and Astronomy, University of Pittsburgh, Pittsburgh, PA 15260, USA}
\begin{abstract}
We develop a framework for deriving Dyson-Schwinger Equations (DSEs) and Bethe-Salpeter Equation (BSE) in QCD at large $N_c$ limit. The starting point is a modified form (with auxiliary fields) of QCD generating functional. This framework provides a natural order-by-order truncation scheme for DSEs and BSE, and the kernels of the equations up to any order are explicitly given. Chiral symmetry (at chiral limit) is preserved in any order truncation, so it exemplifies the symmetry preserving truncation scheme. It provides a method to study DSEs and BSE beyond the Rainbow-Ladder truncation, and is especially useful to study contributions from non-Abelian dynamics (those arise from gluon self-interactions). We also derive the equation for the quark-ghost scattering kernel, and discuss the Slavnov-Taylor identity connecting the quark-gluon vertex, the quark propagator and the quark-ghost scattering kernel.
\end{abstract}
 \maketitle
%\tableofcontents
%%%%%%%%%%%%%%%%%%%%%%%%%%%%%%%%%%%%%%%%%%%%%%%%%%%%%
\section{Introduction}

Standard form of QCD generating functional is the basis for perturbative expansion, while alternate forms may have advantages in nonperturbative studies. For example, In Ref.~\cite{CR}, Cahill and Roberts introduced a bilocal action to study meson properties and bag model. The action was obtained by taking an approximation that only the gluon propagator was retained. So it is not QCD but a model which is usually called Global Color Model  (GCM) (for a review, see Ref.~\cite{Tan}). It is possible to generalize their method to obtain a bilocal action which is equivalent to QCD action, and the GCM action may be served as the leading order of this new bilocal action. In the present paper, we shall employ a bilocal action which was first introduced in Ref.~\cite{WKWX} and was used to study chiral perturbation theory therein. With this action, various problems can be addressed. In this work we generalize the method used in Ref.~\cite{WKWX} by introducing external sources for single quark fields, gluon fields and ghost fields to study Dyson-Schwinger Equations (DSEs) and Bethe-Salpeter Equation (BSE). In this approach, the parameter $N_c$ in the action is extracted and shown explicitly, which is especially beneficial for large $N_c$ expansion. In the present work, we concentrate on large $N_c$ limit although higher-order corrections can be taken into account in principle. We shall derive DSEs and BSE and discuss their truncations. Since all equations are derived from the same generating functional, consistency among these equations are easily maintained when truncations are made. In particular, chiral symmetry (at chiral limit) is preserved automatically in truncated DSEs and BSE up to any order. This is a more elegant approach than truncating the DSEs and the BSE separately and discussing their consistency thereafter.

DSEs and BSE are important tools in low energy QCD and hadron physics. Particularly, meson spectrum can be calculated with BSE combining the gap equation, i.e. the DSE for the quark propagator. It is well known that these equations must be truncated in practical calculations. The most simplest truncation for the gap equation and the meson BSE is the Rainbow-Ladder (RL) truncation~\cite{MR,MT,CM,MRRS,Kra}. In this truncation, the full quark-gluon vertex is replaced with the bare vertex $\gamma^\mu$ (up to a color matrix), and the BSE kernel takes the ladder approximation. Although many achievements have been made based on the RL truncation, recent researches indicated the necessity of going beyond the RL truncation. For instance, the spectra for light scalar and axial-vector mesons calculated under the RL truncation are relatively poor compared to those for light pseudoscalar and vector mesons~\cite{BQRTT,WCT,Maris1}. The reason is believed that in the pseudoscalar and vector channels the higher-order corrections to the RL truncation largely cancel, while in the scalar and axial-vector channels the corrections do not typically cancel~\cite{Rob,BCPQR}. So, recent years, a lot of efforts were made aiming at going beyond the RL truncation in DSEs and BSE~\cite{BDTR,FW,WF,AW,RHK,AFLS,FNW}.

Chiral symmetry breaking plays an important role in the low energy QCD, so the truncated DSEs and BSE should reflect this feature properly. It is realized that chiral symmetry imposes an important connection between the integration kernel of the gap equation and the BSE kernel. This connection guarantees that the pion state as a solution of the BSE for a quark-antiquark pair is automatically the Goldstone particle when chiral symmetry is spontaneously broken (in the chiral limit)~\cite{Munczek,BRS}. In this respect, the symmetry preserving truncation scheme is proposed, which requires the approximation made in the BSE kernel must be in consistent with the truncation made in the gap equation such that chiral symmetry is preserved in the chiral limit~\cite{BRS}. In this scheme, the well-known RL truncation is just its lowest-order truncation.

To go beyond RL truncation, one can make use of the DSE for the quark-gluon vertex (QGV). This equation explicitly shows that how the strong interaction makes corrections to the bare quark-gluon vertex. At one loop level, keeping all the propagators dressed, the QGV has two triangle Feynman diagrams contributions, which are usually served as the next-to-leading-order correction. One may continue to consider higher-loop contributions, and improve the truncation gradually. This method of going beyond the RL truncation has a benefit that the effects of QCD dynamics in the quark condensate and physical observables can be directly tested. Especially, the gluon self-interactions, which are typical non-Abelian dynamics, can be directly tested~\cite{FW,FUW}. 

Another way of going beyond the RL truncation makes use of the Slavnov-Taylor identity (STI) of the QGV. This STI relates the QGV to the quark propagator and the quark-ghost scattering kernel, thus it can be used to model the QGV with the other two Green's functions~\cite{FA,AP,Roj}. Because this method escapes from the seemingly endless truncations using the QGV DSE and the STI is automatically fulfilled, it becomes one of the major focuses in the studies beyond the RL truncation. However, due to the appearance of the quark-ghost scattering kernel, the STI of QGV is more complicated than its counterpart in QED, i.e., the Ward-Takahashi identity (WTI) for the fermion-photon vertex. In QED, the fermion-photon vertex may be modeled fully by the fermion propagators, such as the BC vertex~\cite{BC} and the CP vertex~\cite{CP1}, and a closed form of DSEs arises accordingly; while in QCD, the additional Green's function makes the modeling of QGV more complicated and introduces new inputs. So some authors use BC vertex or CP vertex in studying QCD DSEs and sacrifices the non-Abelian feature of the STI. Another problem of this method is that the STI or the WTI in QED only fixes the longitudinal part of the vertex, leaving the transverse part undetermined. In this direction, some efforts have been made~\cite{He,QCLRS}.

Since the study on DSEs and BSE beyond-the-RL truncation is one of the major directions in hadron physics, we develop a framework for studying the quark DSE and the meson BSE in an order-by-order truncation scheme. We shall derive a bilocal form of QCD generating functional first, then make the large $N_c$ expansion. After taking large $N_c$ limit, the DSEs and BSE will be derived and the truncation scheme will be presented. It will be clear that in this truncation scheme the leading-order, i.e. RL truncation is just the Abelian approximation, which means all the higher-order corrections are due to non-Abelian type dynamics of QCD. Thus, large $N_c$ limit extracts the non-Abelian dynamics in the DSEs, and is especially useful for testing the effects of gluon self-interactions. 
On the other hand, from direct analysis of the Feynman diagrams, we know that at large $N_c$ limit, mesons are free particles~\cite{tHooft,Witten}, this will simplify the corresponding BSE. Furthermore, if we reformulate QCD in loop space~\cite{LoopEquation}, DSE for Wilson loops becomes a closed equation, in contrast to present DSEs and BSE which have infinite tower of coupled equation groups and then needs truncation. Although we still do not fully understand why ordinary DSEs at large $N_c$ limit are not closed, while in loop space will close itself, this result appeared in loop space will certainly has some effects on the ordinary DSEs and make large $N_c$ limit different from other kind of approximations of these equations.

In our approach, the higher-order corrections contribute through higher-loop diagrams in the equations. This kind of way of going beyond the RL truncation were studied previously in a number of works as mentioned before. Some of the works concentrated on the next-leading-order corrections without giving explicit forms for higher orders~\cite{FW,WF,FUW,BT}; some of them only considered Abelian-type diagrams and the non-Abelian contributions were absorbed into parameters~\cite{BDTR}. For the former case, because the QGV DSE may be written in different forms~\cite{AFLS}, the truncated gap equations arising from different forms will actually be different in higher orders, which causes an ambiguity. So a systematic way to give a unique order-by-order improvements in the truncated gap equation is called for. Our method presented here provides such a scheme under large $N_c$ limit, in which each order is just an integration of the connected gluon Green's functions in the corresponding order. Moreover, considering large $N_c$ expansion, $1/N_c$ corrections and higher-order contributions can also be taken into account systematically in principle. So eventually, we could have a systematic way to approach the fundamental theory: QCD, in contrast to only provide a model. For the latter case, it is already known that non-Abelian corrections in the next leading-order are dominant compared to Abelian contributions~\cite{BDTR,WF,FUW,BT}. And we shall show that this is true up to any order. So considering non-Abelian type contributions in higher orders is of more significance. Our truncation scheme does exactly this job. In addition, as indicated in Ref.~\cite{BCPQR}, the H-type diagrams in BS kernel are important, because a ladder-like BS kernel with numbers of crossed-box diagrams is insufficient to preserve the Ward-Green-Takahashi identities in general. The H-type diagrams can only originated from non-Abelian diagrams.  

Our framework treats the gauge sector and the fermion sector separately, and its focus is the quark DSE, the QGV DSE and the meson BSE. We also derived the equation for the quark-ghost scattering kernel, then all the Green's functions included in the QGV STI are expressed in the same framework. We hope this could be useful to verify the STI or shed some light on the modeling of QGV using STI. 

The remaining part of this paper is organized as follows. We start with developing the generating functional for deriving the DSEs in Section \uppercase\expandafter{\romannumeral2}. The gap equation and the QGV DSE at large $N_c$ are then derived in Section \uppercase\expandafter{\romannumeral3}. The meson BSE is derived from the same generating functional in Section \uppercase\expandafter{\romannumeral4} and the symmetry preserving truncation scheme is illustrated. Section \uppercase\expandafter{\romannumeral5} is devoted to the discussion on the STI. A summary is followed in Section \uppercase\expandafter{\romannumeral6}.

\section{The generating functional at large $N_c$ limit}
Consider a QCD-type gauge theory with $SU(N_{c})$ local gauge symmetry. Let $A_{\mu}^i~ (i=1,2,\cdots,N_c^2-1)$ be the gauge field, ${\psi}_{\alpha}^{a\eta}$ be , respectively,
fermion fields with color index ${\alpha}~~({\alpha}=1,2,\cdots,N_{c})$, Lorentz spinor index $\eta$, flavor index $a~~(a=1,2,\cdots,N_f)$. For convenience, we simply call ${\psi}_{\alpha}^{a\eta}$ the "quark field" and $A_{\mu}^i$ the "gluon field". Let us introduce local external sources ${\cal I}_i^\mu$ for gauge field $A_{\mu}^i$,
$\bar{I}_{\alpha}^{a\eta}$ for ${\psi}_{\alpha}^{a\eta}$, $I_{\alpha}^{a\eta}$ for $\bar{\psi}_{\alpha}^{a\eta}$, and $J_{\sigma\rho}$ for the composite quark fields
$\bar{\psi}^{\sigma}{\psi^{\rho}}$, where $\sigma$ and $\rho$ are short notations for the spinor and flavor indices. The external source $J$ can be
decomposed into scalar, pseudoscalar, vector, axial-vector and tensor parts
\begin{eqnarray}                              %(1)
J(x)=-s(x) +ip(x)\gamma_5 +v\!\!\! /\;(x) +a\!\!\! /\;(x)\gamma_5+\sigma_{\mu\nu}\bar{t}^{\mu\nu}(x),
\end{eqnarray}
where $s(x)$, $p(x)$, $v_{\mu}(x)$ and $a_{\mu}(x)$  are hermitian matrices, and the quark masses have been absorbed into the definition of $s(x)$. The vector and axial-vector sources $v\!\!\! /\;(x)$, $a\!\!\! /\;(x)$ and $\bar{t}^{\mu\nu}(x)$ are taken to be traceless.

We start from constructing the following generating functional
\begin{eqnarray}                                %(2)
Z[J,{\cal I},\bar{I},I]&=&\int{\cal D}\psi{\cal D}\bar{\psi}{\cal D}A_{\mu}\exp i{\int}d^{4}x\{{\cal L}({\psi},\bar{\psi},A_{\mu})+\bar{\psi}J\psi+{\cal I}_i^{\mu}A_{\mu}^i+\bar{I}\psi+\bar{\psi}I\}\nonumber\\
 &&\hspace*{-2cm}=\int{\cal D}\psi{\cal D}\bar{\psi}\exp\bigg\{i\int d^4 x
\{\bar{\psi}(i\partial\!\!\!/ +J)\psi+\bar{I}\psi+\bar{\psi}I\}\bigg\}\int{\cal D}A_{\mu}{\Delta}_{F}(A_{\mu})\exp\bigg\{ i{\int}d^{4}x\bigg[{\cal L}_{G}(A)
-\frac{1}{2\xi}[F^i(A_{\mu})]^{2}+{\cal I}_i^{\prime\mu}A^i_{\mu}\bigg]\bigg\},~~~~\label{QCDZ}
\end{eqnarray}
where ${\cal L}_{G}(A)=-\frac{1}{4}A^i_{\mu\nu}A^{i\mu\nu}$ is the gluon kinetic energy term, ${\cal I}_i^{\prime\mu}\equiv{\cal I}_i^{\mu}-g\bar{\psi}\frac{\lambda_i}{2}{\gamma}^{\mu}{\psi}$ absorb the quark current (or gauge interaction term) into external source for gauge field, $-\frac{1}{2\xi}[F^i(A_{\mu})]^{2}$ is the gauge-fixing term and ${\Delta}_{F}(A_{\mu})$ is the Fadeev-Popov determinant. The traditional QCD generating functional will be arrived by taking $N_c=3$ and the limit $J(x)\rightarrow -M$ with $M$ the quark mass matrix. We introduce the external source $J$ for composite quark fields to keep the generating functional's potential for further use, and it will not harm current study.

Let us first consider the integration over ${\cal D}A_{\mu}$ for a given configuration of $\psi$ and $\bar{\psi}$, i.e. the current ${\cal I}_i^{\prime\mu}$ serves as an effective external source in the integration over ${\cal D}A_{\mu}$. The result of such an integration can be formally written as
\begin{eqnarray}       ~~~~\label{YMZ}                             %(3)
 &&\int{\cal D}A_{\mu}{\Delta}_{F}(A_{\mu})\exp\bigg\{i{\int}d^{4}x\bigg[{\cal L}_{G}(A)-\frac{1}{2\xi}[F^i(A_{\mu})]^{2}+{\cal I}_i^{\prime\mu}A^i_{\mu}\bigg]\bigg\}\nonumber\\
 &&=\exp\;i\sum^{\infty}_{n=2}{\int}d^{4}x_1\cdots{d^{4}x_n}
\frac{i^n}{n!} G_{\mu_1\cdots\mu_n}^{i_1\cdots i_n}(x_1,\cdots,x_n){\cal I}^{\prime\mu_1}_{i_1}(x_1)\cdots{{\cal I}^{\prime\mu_{n}}_{i_n}(x_n)},
\end{eqnarray}
where $G_{\mu_1\cdots\mu_n}^{i_1\cdots{i_n}}$ is the full connected n-gluon Green's function without inner quark loops. Precisely it is defined as
\begin{eqnarray}
&&i^n G_{\mu_1\cdots\mu_n}^{i_1\cdots i_n}(x_1,\cdots,x_n)\equiv i^n\langle 0|T[A_{\mu_1}^{i_1}(x_1)\cdots A_{\mu_n}^{i_n}(x_n)]|0\rangle_{\mathrm{connected,pure YM}}\nonumber\\
&&=\frac{\delta^n}{\delta{\cal I}^{\mu_1}_{i_1}(x_1)\cdots\delta{\cal I}^{\mu_{n}}_{i_n}(x_n)}
(-i)\ln\int{\cal D}A_{\mu}{\Delta}_{F}(A_{\mu})\exp\bigg\{i{\int}d^{4}x\bigg[{\cal L}_{G}(A)-\frac{1}{2\xi}[F^i(A_{\mu})]^{2}+{\cal I}_i^{\mu}A^i_{\mu}\bigg]\bigg\}\bigg|_{{\cal I}^{\mu}_i(x)=0}
\end{eqnarray}
Note that if the gauge interaction is not non-Abelian but Abelian, then only 2-gluon Green's function (here and later on, we omit ``without inner quark loops" for convenience) is nonzero due to the absence of self interactions among the gauge fields. Hence, we define ``Abelian approximation" as that we only keep the
2-gluon Green's functions in the result and drop out 3-point and more higher points gluon Green's functions.

The source terms in Eq. (\ref{YMZ}) can be written explicitly  as,
\begin{eqnarray}
&&\hspace{-0.5cm}\int{\cal D}A_{\mu}{\Delta}_{F}(A_{\mu})\exp\bigg\{i{\int}d^{4}x\bigg[{\cal L}_{G}(A)-\frac{1}{2\xi}[F^i(A_{\mu})]^{2}+{\cal I}_i^{\prime\mu}A^i_{\mu}\bigg]\bigg\}\nonumber\\
&&\hspace*{-1.0cm}=\exp\;i\sum^{\infty}_{n=2}{\int}d^{4}x_1\cdots{d^{4}x_n}
\frac{i^n}{n!} G_{\mu_1\cdots\mu_n}^{i_1\cdots i_n}(x_1,\cdots,x_n)\nonumber\\
&&\hspace*{-0.5cm}\times\bigg[
[-g\bar{\psi}^{a_1}_{{\alpha}_1}(x_1)(\frac{\lambda_{i_1}}{2})_{\alpha_1\beta_1}\gamma^{\mu_1}{\psi}^{a_1}_{{\beta}_1}(x_1)]\cdots
[-g\bar{\psi}^{a_n}_{{\alpha}_n}(x_n)(\frac{\lambda_{i_n}}{2})_{\alpha_n\beta_n}\gamma^{\mu_n}{\psi}^{a_n}_{\beta_n}(x_n)]\nonumber\\
&&+n{\cal I}^{\mu_1}_{i_1}(x_1)[-g\bar{\psi}^{a_2}_{{\alpha}_2}(x_2)(\frac{\lambda_{i_2}}{2})_{\alpha_2\beta_2}\gamma^{\mu_2}{\psi}^{a_2}_{{\beta}_2}(x_2)]\cdots
[-g\bar{\psi}^{a_n}_{{\alpha}_n}(x_n)(\frac{\lambda_{i_n}}{2})_{\alpha_n\beta_n}\gamma^{\mu_n}{\psi}^{a_n}_{\beta_n}(x_n)]\nonumber\\
&&+n(n-1){\cal I}^{\mu_1}_{i_1}(x_1){\cal I}^{\mu_2}_{i_2}(x_2)[-g\bar{\psi}^{a_3}_{{\alpha}_3}(x_3)(\frac{\lambda_{i_3}}{2})_{\alpha_3\beta_3}\gamma^{\mu_3}{\psi}^{a_3}_{{\beta}_3}(x_3)]\cdots
[-g\bar{\psi}^{a_n}_{{\alpha}_n}(x_n)(\frac{\lambda_{i_n}}{2})_{\alpha_n\beta_n}\gamma^{\mu_n}{\psi}^{a_n}_{\beta_n}(x_n)]\nonumber\\
&&+\cdots+{\cal I}^{\mu_1}_{i_1}(x_1)\cdots{{\cal I}^{\mu_{n}}_{i_n}(x_n)}\bigg],
\end{eqnarray}

By Fierz reordering, we can diagonalize the color indices of the quark fields. For the source independent terms, we have
\begin{equation}
\begin{split}
 &\int d^4x_2\cdots d^4x_n G_{\mu_1\cdots\mu_n}^{i_1\cdots{i_n}} (x_1,\cdots,x_n)[-g\bar{\psi}^{a_1}_{{\alpha}_1}(x_1) (\frac{\lambda_{i_1}}{2})_{\alpha_1\beta_1}\gamma^{\mu_1}
 {\psi}^{a_1}_{{\beta}_1}(x_1)]\cdots[-g\bar{\psi}^{a_n}_{{\alpha}_n}(x_n)(\frac{\lambda_{i_n}}{2})_{\alpha_n\beta_n}\gamma^{\mu_n}{\psi}^{a_n}_{\beta_n}(x_n)]\\
&={\int} d^4x_2\cdots d^4x_nd^{4}x'_1\cdots{d^{4}x'_n}(-1)^ng^{2n-2}\bar{G}^{\sigma_1\cdots\sigma_n}_{\rho_1
\cdots\rho_n}(x_1,x'_1,\cdots,x_n,x'_n)\bar{\psi}^{\sigma_1}_{\alpha_1}(x_1){\psi}^{\rho_1}_{\alpha_1}(x'_1)\cdots\bar{\psi}^{\sigma_n}
 _{\alpha_n}(x_n){\psi}^{\rho_n}_{\alpha_n}(x'_n).
 \end{split}\label{GbarDef}
\end{equation}
where $\bar{G}^{\sigma_1\cdots\sigma_n}_{\rho_1\cdots\rho_n}(x_1,x'_1,\cdots,x_n,x'_n)$ is a extended Green's function containing 2n space-time points.
Using the relations listed in Appendix A, one can check that the extended 2-point Green's function is
\begin{eqnarray}\label{Ext2Gluon}
\overline{G}_{\rho_1\rho_2}^{\sigma_1\sigma_2}(x_1,x_1',x_2,x_2')=-\frac{1}{2}G_{\mu_1\mu_2}(x_1,x_2)[&&(\gamma^{\mu_1})_{\sigma_1\rho_2}(\gamma^{\mu_2})_{\sigma_2\rho_1}
\delta(x_1'-x_2)\delta(x_2'-x_1)+\nonumber\\
&&+\frac{1}{N_c}(\gamma^{\mu_1})_{\sigma_1\rho_1}(\gamma^{\mu_2})_{\sigma_2\rho_2}\delta(x_1'-x_1)\delta(x_2'-x_2)],
\end{eqnarray}
where $\sigma,\rho$'s are combined spinor and flavor indices.

Similarly, we introduce extended Green's functions $\tilde{G}$ for linear source dependent terms, which satisfy
\begin{equation}
\begin{split}
 &\int d^4x_2\cdots d^4x_n G_{\mu_1\cdots\mu_n}^{i_1\cdots{i_n}} (x_1,\cdots,x_n){\cal I}^{\mu_1}_{i_1}(x_1)[-g\bar{\psi}^{a_2}_{{\alpha}_2}(x_2) (\frac{\lambda_{i_2}}{2})_{\alpha_2\beta_2}\gamma^{\mu_2}
 {\psi}^{a_2}_{{\beta}_2}(x_2)]\cdots[-g\bar{\psi}^{a_n}_{{\alpha}_n}(x_n)(\frac{\lambda_{i_n}}{2})_{\alpha_n\beta_n}\gamma^{\mu_n}{\psi}^{a_n}_{\beta_n}(x_n)]\\
&={\int}d^4x_2\cdots d^4x_n d^{4}x'_1\cdots{d^{4}x'_n}(-1)^{n-1}g^{2n-3}{\cal I}^{\mu_1}_{i_1}(x_1)(\lambda_{i_1})_{\alpha\beta}\tilde{G}^{\sigma\sigma_3\cdots\sigma_n}_{\mu_1,\rho\rho_3
\cdots\rho_n}(x_1,x_1',\cdots,x_n,x'_n)\\
&\hspace*{3cm}\times\bar{\psi}^{\sigma}_{\alpha}(x_1'){\psi}^{\rho}_{\beta}(x'_2)
\bar{\psi}^{\sigma_3}_{\alpha_3}(x_3){\psi}^{\rho_3}_{\alpha_3}(x'_3)\cdots\bar{\psi}^{\sigma_n}
 _{\alpha_n}(x_n){\psi}^{\rho_n}_{\alpha_n}(x'_n).\label{GtildeDef}
\end{split}
\end{equation}

Now, Eq. (\ref{QCDZ}) can be written as
\begin{eqnarray}
Z[J,{\cal I},\bar{I},I]&=&\int{\cal D}\psi{\cal D}\bar{\psi}\exp i\bigg\{\int d^{4}x\{
\bar{\psi}(i\partial\!\!\!/+J){\psi}+\bar{I}\psi+\bar{\psi}I\}+\sum^{\infty}_{n=2}{\int}d^{4}x_1\cdots{d^4}x_{n}
d^{4}x_{1}'\cdots{d^4}x_{n}'\bigg[\nonumber\\
&&\times\frac{(-i)^n (g^2)^{n-1}}{n!}\bar{G}^{\sigma_1\cdots\sigma_n}_{\rho_1\cdots\rho_n}(x_1,x'_1,\cdots,x_n,x'_n)\bar{\psi}^{\sigma_1}_{\alpha_1}(x_1)
{\psi}^{\rho_1} _{\alpha_1}(x'_1)\cdots\bar{\psi}^{\sigma_n}_{\alpha_n}(x_n){\psi}^{\rho_n}_{\alpha_n}(x'_n)\label{QCDZ'}\\
&&+\frac{i(-i)^{n-1}g^{2n-3}}{(n-1)!}{\cal I}^{\mu}_i(x_1)(\lambda_i)_{\alpha\beta}\tilde{G}^{\sigma\sigma_3\cdots\sigma_n}_{\mu,\rho\rho_3
\cdots\rho_n}(x_1,x_1',\cdots,x_n,x'_n)\bar{\psi}^{\sigma}_{\alpha}(x_1'){\psi}^{\rho}_{\beta}(x'_2)
\bar{\psi}^{\sigma_3}_{\alpha_3}(x_3){\psi}^{\rho_3}_{\alpha_3}(x'_3)\cdots\bar{\psi}^{\sigma_n}
 _{\alpha_n}(x_n){\psi}^{\rho_n}_{\alpha_n}(x'_n)\nonumber\\
 &&+O({\cal I}^2)\bigg]\bigg\}.\nonumber
\end{eqnarray}

In order to integrate out the quark fields $\psi$ and $\bar{\psi}$, we introduce a bilocal auxiliary field $\Phi^{(a\eta)(b\zeta)}(x,x')$ by inserting into (\ref{QCDZ'}) the following constant
\begin{eqnarray}                            %(6)
\int {\cal D}\Phi~\delta \bigg(N_c\Phi^{(a\eta)(b\zeta)}(x,x')-
\bar{\psi}^{a\eta}_\alpha(x)\psi^{b\zeta}_\alpha(x')\bigg).\label{Phi}
\end{eqnarray}
We see from (6) that the bilocal auxiliary field $\Phi^{(a\eta)(b\zeta)}(x,x')$ embodies the bilocal composite fields $\bar{\psi}^{a\eta}_\alpha(x)\psi^{b\zeta}_\alpha(x')$ which reflects
the meson fields. Inserting (\ref{Phi}) into (\ref{QCDZ'}) we get
\begin{eqnarray}                         %(5)
Z[J,{\cal I},\bar{I},I]&=&\int{\cal D}\psi{\cal D}\bar{\psi}{\cal D}\Phi\delta\bigg(N_c \Phi^{(a\eta)(b\zeta)}(x,x')-\bar{\psi}^{a\eta}_{\alpha}(x)
\psi^{b\zeta}_{\alpha}(x')\bigg)\exp i\bigg\{\int d^{4}x\{
\bar{\psi}(i\partial\!\!\!/+J){\psi}+\bar{I}\psi+\bar{\psi}I\}\nonumber\\
&&\hspace*{-2cm}+\sum^{\infty}_{n=2}{\int}d^{4}x_1\cdots{d^4}x_{n}
d^{4}x_{1}'\cdots{d^4}x_{n}'\bigg[N_c\frac{(-i)^n (N_cg^2)^{n-1}}{n!}\bar{G}^{\sigma_1\cdots\sigma_n}_{\rho_1\cdots\rho_n}(x_1,x'_1,\cdots,x_n,x'_n)
\Phi^{\sigma_1\rho_1}(x_1 ,x'_1)\cdots\Phi^{\sigma_n\rho_n}(x_n ,x'_n)~~~\label{defPhi}\\
&&\hspace*{-2cm}+\frac{i(-i)^{n-1}g^{2n-3}N_c^{n-2}}{(n-1)!}{\cal I}^{\mu}_i(x_1)(\lambda_i)_{\alpha\beta}\tilde{G}^{\sigma\sigma_3\cdots\sigma_n}_{\mu,\rho\rho_3
\cdots\rho_n}(x_1,x_1',\cdots,x_n,x'_n)\bar{\psi}^{\sigma}_{\alpha}(x_1'){\psi}^{\rho}_{\beta}(x'_2)
\Phi^{\sigma_3\rho_3}(x_3 ,x'_3)\cdots\Phi^{\sigma_n\rho_n}(x_n ,x'_n)+O({\cal I}^2)\bigg]\bigg\}.\nonumber
\end{eqnarray}
The $\delta$-function in (\ref{defPhi}) can be further expressed in the Fourier representation
\begin{eqnarray}
\delta \bigg(N_c\Phi(x,x')-\bar{\psi}(x)\psi(x')\bigg)\sim\int{\cal D}\Pi e^{i\int d^4xd^4x'\Pi(x,x')\cdot
\big(N_c\Phi(x,x')-\bar{\psi}(x)\psi(x')\big)}.
\nonumber
\end{eqnarray}
The generating functional then becomes
\begin{eqnarray}                         %(5)
Z[J,{\cal I},\bar{I},I]&=&\int{\cal D}\psi{\cal D}\bar{\psi}{\cal D}\Phi{\cal D}\Pi\exp i\bigg\{\int d^{4}x\{
\bar{\psi}(i\partial\!\!\!/+J-\Pi){\psi}+\bar{I}\psi+\bar{\psi}I\}+\int d^{4}xd^{4}x'N_c \Phi^{\sigma\rho}(x,x')\Pi^{\sigma\rho}(x,x')\nonumber\\
&&\hspace*{-1cm}+\sum^{\infty}_{n=2}{\int}d^{4}x_1\cdots{d^4}x_{n}
d^{4}x_{1}'\cdots{d^4}x_{n}'N_c\frac{(-i)^n (N_cg^2)^{n-1}}{n!}\bar{G}^{\sigma_1\cdots\sigma_n}_{\rho_1\cdots\rho_n}(x_1,x'_1,\cdots,x_n,x'_n)
\Phi^{\sigma_1\rho_1}(x_1 ,x'_1)\cdots\Phi^{\sigma_n\rho_n}(x_n ,x'_n)\nonumber\\
&&\hspace*{-1cm}+\int d^4x_1d^4x_1'd^4x_2'\bar{\psi}^{\sigma}_{\alpha}(x_1')\tilde{\cal I}^{\mu}_{\alpha\beta}(x_1)\Delta^{\sigma\rho}_{\Phi,\mu}(x_1,x_1',x_2'){\psi}^{\rho}_{\beta}(x'_2)
+O({\cal I}^2)\bigg\}.~~~\label{defPhi1}
\end{eqnarray}
where $\tilde{\cal I}^{\mu}_{\alpha\beta}(x_1)\equiv{\cal I}^{\mu}_i(\lambda_i)_{\alpha\beta}$ and
\begin{eqnarray}
\Delta^{\sigma\rho}_{\Phi,\mu}(x_1,x_1',x_2')&\equiv&\int d^4x_2\cdots d^4x_nd^4x_3'\cdots d^4x_n'
\frac{i(-i)^{n-1}g^{2n-3}N_c^{n-2}}{(n-1)!}\tilde{G}^{\sigma\sigma_3\cdots\sigma_n}_{\mu,\rho\rho_3\cdots\rho_n}(x_1,x_1',\cdots,x_n,x'_n)\nonumber\\
&&\times\Phi^{\sigma_3\rho_3}(x_3 ,x'_3)\cdots\Phi^{\sigma_n\rho_n}(x_n ,x'_n)
\end{eqnarray}
Integrating out the $\psi$ and $\bar{\psi}$ fields leads us to
\begin{eqnarray}
Z[J,{\cal I},\bar{I},I]&=&\int{\cal D}\Phi{\cal D}\Pi
\exp i\bigg\{ -i{\rm Tr}'\ln[i\partial\!\!\!/+J-\Pi+\tilde{\cal I}\Delta_\Phi]-\bar{I}[i\partial\!\!\!/+J-\Pi+\tilde{\cal I}\Delta_\Phi]^{-1}I\nonumber\\
&&+\int d^{4}xd^{4}x'N_c \Phi^{\sigma\rho}(x,x')\Pi^{\sigma\rho}(x,x')+N_c \sum^{\infty}_{n=2}{\int}d^{4}x_1\cdots{d^4}x_{n}
d^{4}x_{1}'\cdots{d^4}x_{n}'\frac{(-i)^{n}(N_c g^2)^{n-1}}{n!}\nonumber\\
&&\times\bar{G}^{\sigma_1\cdots\sigma_n}_{\rho_1\cdots\rho_n}(x_1,x'_1,\cdots,x_n,x'_n)\Phi^{\sigma_1\rho_1}(x_1 ,x'_1)\cdots\Phi^{\sigma_n\rho_n}(x_n ,x'_n)+O({\cal I}^2) \bigg\}\label{W00},
\end{eqnarray}
where ${\rm Tr}'$ is the functional trace with respect to the space-time, color, spinor and flavor indices. Expanding the first two terms on the exponential with respect to $\tilde{\cal I}$, we arrive at
\begin{eqnarray}
Z[J,{\cal I},\bar{I},I]&=&\int{\cal D}\Phi{\cal D}\Pi
\exp i\bigg\{ -iN_c{\rm Tr}\ln[i\partial\!\!\!/+J-\Pi]-\bar{I}[i\partial\!\!\!/+J-\Pi]^{-1}I+\bar{I}[i\partial\!\!\!/+J-\Pi]^{-1}\tilde{\cal I}\Delta_\Phi[i\partial\!\!\!/+J-\Pi]^{-1}I\nonumber\\
&&+\int d^{4}xd^{4}x'N_c \Phi^{\sigma\rho}(x,x')\Pi^{\sigma\rho}(x,x')+N_c \sum^{\infty}_{n=2}{\int}d^{4}x_1\cdots{d^4}x_{n}
d^{4}x_{1}'\cdots{d^4}x_{n}'\frac{(-i)^{n}(N_c g^2)^{n-1}}{n!}\nonumber\\
&&\times\bar{G}^{\sigma_1\cdots\sigma_n}_{\rho_1\cdots\rho_n}(x_1,x'_1,\cdots,x_n,x'_n)\Phi^{\sigma_1\rho_1}(x_1 ,x'_1)\cdots\Phi^{\sigma_n\rho_n}(x_n ,x'_n)+O({\cal I}^2)\bigg\}.\label{W01}
\end{eqnarray}
The difference of Tr and Tr' is that Tr do not include trace of color index.

Taking large $N_c$ limit, we obtain
\begin{eqnarray}
Z'[J,{\cal I},\bar{I},I]&\equiv&\lim_{N_c\rightarrow\infty}Z[J,{\cal I},\bar{I},I]\nonumber\\
&=&\mbox{const}\times
\exp i\bigg\{ -iN_c{\rm Tr}\ln[i\partial\!\!\!/+J-\Pi_c]-\bar{I}[i\partial\!\!\!/+J-\Pi_c]^{-1}I\nonumber\\
&&+\bar{I}[i\partial\!\!\!/+J-\Pi_c]^{-1}\tilde{\cal I}\Delta_{\Phi_c}[i\partial\!\!\!/+J-\Pi_c]^{-1}I\nonumber\\
&&+\int d^{4}xd^{4}x'N_c \Phi_c^{\sigma\rho}(x,x')\Pi_c^{\sigma\rho}(x,x')+N_c \sum^{\infty}_{n=2}{\int}d^{4}x_1\cdots{d^4}x_{n}
d^{4}x_{1}'\cdots{d^4}x_{n}'\frac{(-i)^{n}(N_c g^2)^{n-1}}{n!}\nonumber\\
&&\times\bar{G}^{\sigma_1\cdots\sigma_n}_{\rho_1\cdots\rho_n}(x_1,x'_1,\cdots,x_n,x'_n)\Phi_c^{\sigma_1\rho_1}(x_1 ,x'_1)\cdots\Phi_c^{\sigma_n\rho_n}(x_n ,x'_n)+O({\cal I}^2)\bigg\},
\label{W02}
\end{eqnarray}
where $O_c$ is the expectation value of $O$, i.e. $O_c\equiv\frac{\int {\cal D}\Phi{\cal D}\Pi O e^{iS}}{\int {\cal D}\Phi{\cal D}\Pi e^{iS}}$, and $iS$ is the exponent in Eq. (\ref{W01}).
In this paper, whenever large $N_c$ limit is taken, it is implied that only leading terms in $1/N_c$ expansion are retained in the n-gluon Green's functions $\bar{G}$ and $\tilde{G}$.

\section{Deriving the gap equation and the QGV DSE}
The exponential of the generating functional given in Eq. (\ref{W02}) is also the effective action for $\Pi_c$ and $\Phi_c$, so $\Pi_c$ and $\Phi_c$ satisfy stationery equations, which give
\begin{eqnarray}
\frac{-i\delta\ln Z'[J,{\cal I},\bar{I},I]}{\delta\Pi_c^{\sigma\rho}(x,y)}
&=&N_c\bigg[i[i\partial\!\!\!/+J-\Pi_c]^{-1,\rho,\sigma}(y,x)+\Phi_c^{\sigma\rho}(x,y)\bigg]+\bigg[[i\partial\!\!\!/+J-\Pi_c]^{-1}I\bar{I}[i\partial\!\!\!/+J-\Pi_c]^{-1}\bigg]^{\sigma\rho}(x,y)\nonumber\\
&&-\bigg[[i\partial\!\!\!/+J-\Pi_c]^{-1}\tilde{\cal I}\Delta_{\Phi_c}[i\partial\!\!\!/+J-\Pi_c]^{-1}I\bar{I}[i\partial\!\!\!/+J-\Pi_c]^{-1}\bigg]^{\sigma\rho}(x,y)\nonumber\\
&&-\bigg[[i\partial\!\!\!/+J-\Pi_c]^{-1}I\bar{I}[i\partial\!\!\!/+J-\Pi_c]^{-1}\tilde{\cal I}\Delta_{\Phi_c}[i\partial\!\!\!/+J-\Pi_c]^{-1}\bigg]^{\sigma\rho}(x,y)\nonumber\\
&=&0,\label{Pidef}
\end{eqnarray}
and
\begin{eqnarray}
\frac{-i\delta\ln Z'[J,{\cal I},\bar{I},I]}{\delta\Phi_c^{\sigma\rho}(x,y)}
&=&N_c\bigg[\Pi_c^{\sigma\rho}(x,y)+\sum^{\infty}_{n=2}{\int}d^{4}x_2\cdots{d^4}x_{n}
d^{4}x_{2}'\cdots{d^4}x_{n}'\frac{(-i)^{n}(N_c g^2)^{n-1}}{(n-1)!}\bar{G}^{\sigma\sigma_2\cdots\sigma_n}_{\rho\rho_2\cdots\rho_n}(x,y,x_2,x'_2,\cdots,x_n,x'_n)\nonumber\\
&&\times\Phi_c^{\sigma_2\rho_2}(x_2 ,x'_2)\cdots\Phi_c^{\sigma_n\rho_n}(x_n ,x'_n)\bigg]+\bar{I}[i\partial\!\!\!/+J-\Pi_c]^{-1}\tilde{\cal I}\frac{\delta\Delta_{\Phi_c}}{\delta\Phi_c^{\sigma\rho}(x,y)}[i\partial\!\!\!/+J-\Pi_c]^{-1}I \nonumber\\
&=&0,\label{Phidef}
\end{eqnarray}
up to $O({\cal I}^1)$ order. Switching off all external sources, the above equations are reduced to
\begin{eqnarray}
\Phi_c^{\sigma\rho}(x,y)&=&-i[i\partial\!\!\!/-M-\Pi_c]^{-1,\rho,\sigma}(y,x),\label{SelfEdef}\\
\Pi_c^{\sigma\rho}(x,y)&=&-\sum^{\infty}_{n=2}{\int}d^{4}x_2\cdots{d^4}x_{n}
d^{4}x_{2}'\cdots{d^4}x_{n}'\frac{(-i)^{n}(N_c g^2)^{n-1}}{(n-1)!}\bar{G}^{\sigma\sigma_2\cdots\sigma_n}_{\rho\rho_2\cdots\rho_n}(x,y,x_2,x'_2,\cdots,x_n,x'_n)\nonumber\\
&&\times\Phi_c^{\sigma_2\rho_2}(x_2 ,x'_2)\cdots\Phi_c^{\sigma_n\rho_n}(x_n ,x'_n).\label{SDeq}
\end{eqnarray}
The first equation says that $i\Phi_c$ is just the quark propagator and $\Pi_c$ is the quark self energy (up to a color factor), which can be checked by evaluating the functional derivative with respect to $I$ and $\bar{I}$:
\begin{eqnarray}
{\langle 0|T[\psi^{\rho}_\beta(y)\bar{\psi}^{\sigma}_{\alpha}(x)]|0\rangle}_c&=&i\frac{\delta^2\ln Z'[J,{\cal I},\bar{I},I]}{\delta\bar{I}^{\rho}_{\beta}(y)\delta {(-I^{\sigma}_{\alpha}(x))}}\bigg|_{J=-M;{\cal I}=\bar{I}=I=0}\nonumber\\
&=&\delta_{\alpha\beta}[i\partial\!\!\!/-M-\Pi_c]^{-1,\rho\sigma}(y,x).
\end{eqnarray}

In deriving this equation, one should notice that due to Eq. (\ref{Pidef}) and Eq. (\ref{Phidef}), only those terms with external sources explicitly would contribute, i.e.
\begin{equation}\label{FuncDer}
\frac{\delta \ln Z'}{ \delta {(-I^{\sigma}_{\alpha}(x))}}=\frac{\delta \ln Z'}{\delta \Phi_c}\frac{\delta \Phi_c}{\delta {(-I^{\sigma}_{\alpha}(x))}}+\frac{\delta \ln Z'}{\delta \Pi_c}\frac{\delta \Pi_c}{\delta {(-I^{\sigma}_{\alpha}(x))}}+\frac{\partial \ln Z'}{ \partial {(-I^{\sigma}_{\alpha}(x))}}=\frac{\partial \ln Z'}{ \partial {(-I^{\sigma}_{\alpha}(x))}}.
\end{equation}

Substituting Eq. (\ref{SelfEdef}) into Eq. (\ref{SDeq}), we obtain
\begin{eqnarray}
[-i\Phi_c^{-1}-i\partial\!\!\!/+M]^{\rho\sigma}(y,x)&=&\sum^{\infty}_{n=2}{\int}d^{4}x_2\cdots{d^4}x_{n}
d^{4}x_{2}'\cdots{d^4}x_{n}'\frac{(-i)^{n}(N_c g^2)^{n-1}}{(n-1)!}\bar{G}^{\sigma\sigma_2\cdots\sigma_n}_{\rho\rho_2\cdots\rho_n}(x,y,x_2,x'_2,\cdots,x_n,x'_n)\nonumber\\
&&\times\Phi_c^{\sigma_2\rho_2}(x_2 ,x'_2)\cdots\Phi_c^{\sigma_n\rho_n}(x_n ,x'_n).\label{SDeq-1}
\end{eqnarray}
Then we recognize that this is just the Dyson-Schwinger equation for quark propagator.

Using the generating functional, we can also derive the quark-gluon 3-point Green's function:
\begin{eqnarray}
&&\langle 0|T[\psi^{\rho}_\beta(y)\bar{\psi}^{\sigma}_{\alpha}(x)A_\mu^i(z)]|0\rangle_c=\frac{\delta^3\ln Z'[J,{\cal I},\bar{I},I]}{\delta\bar{I}^{\rho}_{\beta}(y)\delta (-I^{\sigma}_{\alpha}(x))\delta{\cal I}^\mu_i(z)}\bigg|_{J=-M;{\cal I}=\bar{I}=I=0}\nonumber\\
&=&i(\lambda_i)_{\beta\alpha}\int d^4z_1^4d^4z_2^4[i\partial\!\!\!/-M-\Pi_c]^{-1,\rho\rho'}(y,z_1)\Delta_{\Phi_c,\mu}^{\rho'\sigma'}(z,z_1,z_2)[i\partial\!\!\!/-M-\Pi_c]^{-1,\sigma'\sigma}(z,x).\label{QuarkGluonVertex-1}
\end{eqnarray}

Quark-gluon 3-point Green's function relates to
the quark-gluon vertex $\Gamma_{\mu;\alpha\beta}^i$ as
\begin{eqnarray}
\langle 0|T[\psi^{\rho}_\beta(y)\bar{\psi}^{\sigma}_{\alpha}(x)A_\mu^i(z)]|0\rangle_c
&=&i\int d^4x'd^4y'd^4z'~G_{\mu\nu}^{ij}(z,z')\Phi_c^{\rho\rho'}(y,y')\Gamma^{\nu,\rho'\sigma'}_{j,\beta\alpha}(z',y',x')\Phi_c^{\sigma'\sigma}(x',x).\label{QGGreen}
 \end{eqnarray}
Combining Eq. (\ref{QGGreen}) with Eq.(\ref{QuarkGluonVertex-1}) and  Eq.(\ref{SelfEdef}), we obtain
\begin{eqnarray}
  -\int d^4z'~G_{\mu\nu}(z,z')\Gamma^{\nu,\rho'\sigma'}_{i,\beta\alpha}(z',y',x')&=&(\lambda_i)_{\beta\alpha}\Delta_{\Phi_c,\mu}^{\rho'\sigma'}(z,y',x')\nonumber\\
  &=&(\lambda_i)_{\beta\alpha}\sum^{\infty}_{n=2}\int d^4x_2\cdots d^4x_nd^4x_3'\cdots d^4x_n'
\frac{i(-i)^{n-1}g^{2n-3}N_c^{n-2}}{(n-1)!}\label{QuarkGluonVertex-2}\\
&&\times\tilde{G}^{\rho'\sigma_3\cdots\sigma_n}_{\mu,\sigma'\rho_3\cdots\rho_n}(z,y',x_2,x',x_3,x_3'\cdots,x_n,x'_n)\Phi_c^{\sigma_3\rho_3}(x_3 ,x'_3)\cdots\Phi_c^{\sigma_n\rho_n}(x_n ,x'_n).\nonumber
\end{eqnarray}
We have derived the expression of the QGV at large $N_c$ limit, then the standard form of the gap equation can be written as
\begin{eqnarray}
[-i\Phi_c^{-1}-i\partial\!\!\!/+M]^{\rho\sigma}(x,z)\delta_{\alpha\beta}&=&-g(\frac{\lambda_i}{2})_{\beta\beta'}\gamma^\mu_{\sigma\sigma'}\int d^4y'd^4z'G_{\mu\nu}(z,z')\Phi_c^{\rho'\sigma'}(y',z)\Gamma^{\nu,\rho'\rho}_{i,\beta'\alpha}(z',y',x). \label{SDeqStan}
\end{eqnarray}
Using the Eqs. (\ref{GbarDef}) and (\ref{GtildeDef}), we find the relation between $\tilde{G}$ and $\bar{G}$:
\begin{equation}
\begin{split}
&\delta(\bar{x}_1-x_2)\bar{G}^{\sigma_1\cdots\sigma_n}_{\rho_1
\cdots\rho_n}(x_1,x'_1,\cdots,x_n,x'_n)\\
&=\int d^4\bar{x}_2\bigg\{-\delta(x'_1-\bar{x}_2)\delta(x'_2-x_1)\tilde{G}^{\sigma_2\sigma_3\cdots\sigma_n}_{\mu_1,\rho_1\rho_3
\cdots\rho_n}(x_1,\bar{x}_1,x_2,\bar{x}_2,\cdots,x_n,x'_n)\gamma^{\mu_1}_{\sigma_1\rho_2}\\
 &\hspace*{1cm}-\frac{1}{N_c}\delta(x'_1-x_1)\delta(\bar{x}_2-x'_2)\tilde{G}^{\sigma_2\sigma_3\cdots\sigma_n}_{\mu_1,\rho_2\rho_3
\cdots\rho_n}(x_1,\bar{x}_1,x_2,\bar{x}_2,\cdots,x_n,x'_n)\gamma^{\mu_1}_{\sigma_1\rho_1}\bigg\}.
\end{split}\label{GbarvsGtilde}
\end{equation}
With this relation, one can check that Eq. (\ref{SDeq-1}) and Eq. (\ref{SDeqStan}) are indeed the same.

We have derived the gap equation and the DSE for QGV in our framework, now we can discuss truncations for these equations. As a benefit, truncations can be made either through the QGV DSE and the gap equation directly or through the generating functional $-i\ln Z'$. Truncating the generating functional guarantees all the equations derived from it are in consistent with each others, and any linearly represented symmetry conserved in the truncated generating functional are also conserved by these equations. The last term on the exponential of Eq. (\ref{W02}) is an infinite summation of integrations. Recalling that each extended 2n-point Green's function corresponds to the connected n-gluon Green's function (without quark loops), we find that this summation is an expansion with respect to the number of external legs of the connected gluon Green's functions. So a natural way to approximate the generating functional is to keep finite orders of this expansion. (Due to Eq. (\ref{GbarvsGtilde}), the $\Delta_{\Phi}$ term should be approximated accordingly.)  The lowest order is to keep only the 2-gluon Green's function, i.e. the Abelian approximation. Making use of Eq. (\ref{QuarkGluonVertex-2}) and Eq. (\ref{GbarvsGtilde}), one can find that
\begin{eqnarray}
  \Gamma^{\nu,\rho'\sigma'}_{i,\beta\alpha}(z',y',x')\stackrel{\mathrm{\tiny large~N_c;~Abelian~approximation}}{===================}
  -g(\frac{\lambda_i}{2})_{\beta\alpha}\gamma_{\rho'\sigma'}^{\nu}\delta(z'-y')\delta(z'-x').
\end{eqnarray}
Therefore Abelian approximation at large $N_c$ limit is just the well-known rainbow approximation in which the quark-gluon
vertex takes its bare form.
The next-to-leading order is to keep up to 3-gluon Green's functions in the expansion, which already gives corrections to the bare QGV. Continue to add higher-point gluon Green's functions, then we have an order-by-order truncation scheme.  All the correction terms to the bare QGV are due to non-Abelian type interactions, so it is useful for testing non-Abelian dynamics in a beyond rainbow approximation.

In this framework outlined before, the gauge sector and the fermion sector are treated differently. Pure gluon Green's functions are already introduced in the generating functional. It is convenient to derive any-point quark Green's functions in our method, but we can only derive the equations for 1-gluon-plus-n-quark Green's functions, such as the quark-gluon 3-point Green's function discussed before. The reason is that our current form of the generating functional does not give explicit form for terms including 2 and higher gluon external sources ${\cal I}$. The positive side of this method is that one can concentrate on the fermion sector and treat the gauge sector as inputs which can be extracted from studies on the corresponding pure Yang-Mills theory.

\section{The meson Bethe-Salpeter Equation and chiral symmetry preserving truncations}
We have shown the DSEs derivation and their truncations at large $N_c$ limit in our framework, now we turn to discuss the meson Bethe-Salpeter equation. In order to derive the meson BSE, we derive the 4-point quark Green's function first. Noticing Eq. (\ref{FuncDer}), one can directly have
\begin{eqnarray}
&&\langle 0|T[\psi^{\rho_1}_{\beta_1}(y_1)\bar{\psi}^{\sigma_1}_{\alpha_1}(x_1)\psi^{\rho_2}_{\beta_2}(y_2)\bar{\psi}^{\sigma_2}_{\alpha_2}(x_2)]|0\rangle_c=-i\frac{\delta^4\ln Z'[J,{\cal I},\bar{I},I]}{\delta\bar{I}^{\rho_1}_{\beta_1}(y_1)\delta I^{\sigma_1}_{\alpha_1}(x_1)\delta\bar{I}^{\rho_2}_{\beta_2}(y_2)\delta I^{\sigma_2}_{\alpha_2}(x_2)}\bigg|_{\mathrm{external~sources~vanishing~(e.s.v.)}}\nonumber\\
&&{=\int d^4x'd^4y'\bigg\{\delta_{\alpha_2\beta_2}[i\partial\!\!\!/-M-\Pi_c]^{-1,\rho_2\sigma'}(y_2,x')\frac{\delta\Pi_c^{\sigma'\rho'}(x',y')}{\delta\bar{I}^{\rho_1}_{\beta_1}(y_1)
\delta I^{\sigma_1}_{\alpha_1}(x_1)}\bigg|_{\mathrm{e.s.v.}}[i\partial\!\!\!/-M-\Pi_c]^{-1,\rho'\sigma_2}(y',x_2)}\nonumber\\
&&{\hspace*{2cm}-\delta_{\alpha_2\beta_1}[i\partial\!\!\!/-M-\Pi_c]^{-1,\rho_1\sigma'}(y_1,x')\frac{\delta\Pi_c^{\sigma'\rho'}(x',y')}{\delta\bar{I}^{\rho_2}_{\beta_2}(y_2)
\delta I^{\sigma_1}_{\alpha_1}(x_1)}\bigg|_{\mathrm{e.s.v.}}[i\partial\!\!\!/-M-\Pi_c]^{-1,\rho'\sigma_2}(y',x_2)}\bigg\},
\label{fourquarkGF}
\end{eqnarray}
where we have used equations:
\begin{eqnarray}
\frac{\delta\Pi_c^{\sigma\rho}(x,y)}{\delta{I}^{\sigma'}_\alpha(z)}\bigg|_{\mathrm{e.s.v.}}=\frac{\delta\Pi_c^{\sigma\rho}(x,y)}{\delta{\bar{I}}^{\sigma'}_\alpha(z)}\bigg|_{\mathrm{e.s.v.}}=0
\end{eqnarray}

For convenience, we reduce the notation according to
\begin{eqnarray}
\sigma_i\equiv(\sigma_i,x_i),&&\rho_i\equiv(\rho_i,y_i)\\
\sigma'_i\equiv(\sigma'_i,z_i),&&\rho'_i\equiv(\rho'_i,z'_i)\\
\frac{\delta\Pi_c^{\sigma\rho}(x,y)}{\delta\bar{I}^{\rho_1}_{\beta_1}(y_1)
\delta I^{\sigma_1}_{\alpha_1}(x_1)}\bigg|_{\mathrm{e.s.v.}}&\equiv&\delta\Pi^{\sigma\rho}_{c\rho_1\sigma_1,\beta_1\alpha_1}\\
\frac{\delta\Phi_c^{\sigma\rho}(x,y)}{\delta\bar{I}^{\rho_1}_{\beta_1}(y_1)
\delta I^{\sigma_1}_{\alpha_1}(x_1)}\bigg|_{\mathrm{e.s.v.}}&\equiv&\delta\Phi^{\sigma\rho}_{c\rho_1\sigma_1,\beta_1\alpha_1}.
\end{eqnarray}
With these simplified notations, Eq. (\ref{fourquarkGF}) can be rewritten as
\begin{eqnarray}
\langle 0|T[\psi^{\rho_1}_{\beta_1}\bar{\psi}^{\sigma_1}_{\alpha_1}\psi^{\rho_2}_{\beta_2}\bar{\psi}^{\sigma_2}_{\alpha_2}]|0\rangle_c
&=&\delta_{\alpha_2\beta_2}[i\partial\!\!\!/-M-\Pi_c]^{-1,\rho_2,\sigma'}\delta\Pi^{\sigma'\rho'}_{c\rho_1\sigma_1,\beta_1\alpha_1}[i\partial\!\!\!/-M-\Pi_c]^{-1,\rho',\sigma_2}\nonumber\\
&&-\delta_{\alpha_2\beta_1}[i\partial\!\!\!/-M-\Pi_c]^{-1,\rho_1,\sigma'}\delta\Pi^{\sigma'\rho'}_{c\rho_2\sigma_1,\beta_2\alpha_1}[i\partial\!\!\!/-M-\Pi_c]^{-1,\rho',\sigma_2}.\label{fourquarkGFSim}
\end{eqnarray}
The exact 4-point quark Green's function (without taking large $N_c$ limit) satisfies an inhomogeneous Bethe-Salpeter equation, however, the l.h.s. of Eq. (\ref{fourquarkGFSim}) is not the corresponding Green's function at large $N_c$ limit, because it has 4 free color indices. Only after we extract the colorless part (for instance, timing $\delta_{\alpha_1\beta_1}\delta_{\alpha_2\beta_2}$ then summing over all the color indices), we then obtain a meaningful Green's function under large $N_c$ limit.
This colorless part of the Green's function should satisfy a inhomogeneous Bethe-Salpeter Equation. Since meson only appear in color singlet channels, this is sufficient to derive the homogeneous BSE for mesons.

We may proceed by timing $\delta_{\alpha_1\beta_1}\delta_{\alpha_2\beta_2}$ then summing over all the color indices in Eq. (\ref{fourquarkGFSim}), however, there is a simpler way for our purpose once we notice the relationship between the $\delta\Phi^{\sigma\rho}_{\rho_1\sigma_1,\beta_1\alpha_1}$ and the 4-point quark Green's function.
We have
\begin{equation}
\Phi_c^{\sigma\rho}[I,\bar{I}]=\frac{-\delta_{\alpha\beta}}{N_c}\langle 0|T[\psi^\rho_\beta\bar{\psi}^\sigma_\alpha] |0\rangle_f=\frac{-\delta_{\alpha\beta}}{N_c}\frac{\int [d\phi]\psi^\rho_\beta\bar{\psi}^\sigma_\alpha e^{iS[\phi,I,\bar{I}]}}{\int [d\phi]e^{iS[\phi,I,\bar{I}]}},
\end{equation}
where $\langle\cdots\rangle_f$ denotes the full Green's function which is different from the connected Green's function; $S[\phi,I,\bar{I}]$ is the traditional QCD action with $\phi$ denoting all the basic fields in the QCD Lagrangian (we have omitted the irrelevant external sources). Then
\begin{eqnarray}
\frac{\delta\Phi_c^{\sigma\rho}}{\delta\bar{I}^{\rho_1}_{\beta_1}\delta I^{\sigma_1}_{\alpha_1}}\bigg|_{\bar{I}=I=0}&=&\frac{-\delta_{\alpha\beta}}{N_c}\frac{\delta^2}{\delta\bar{I}^{\rho_1}_{\beta_1}\delta I^{\sigma_1}_{\alpha_1}}\left[\frac{\int [d\phi]\psi^\rho_\beta\bar{\psi}^\sigma_\alpha e^{iS[\phi,I,\bar{I}]}}{\int [d\phi]e^{iS[\phi,I,\bar{I}]}}\right]\bigg|_{\bar{I}=I=0}\nonumber\\
&=&\frac{-\delta_{\alpha\beta}}{N_c}\left\{\frac{1}{Z[0]}\int [d\phi]\psi^\rho_\beta\bar{\psi}^\sigma_\alpha i\psi^{\rho_1}_{\beta_1}(-i\bar{\psi}^{\sigma_1}_{\alpha_1}) e^{iS[\phi]}-\frac{\int [d\phi]\psi^\rho_\beta\bar{\psi}^\sigma_\alpha e^{iS[\phi]}}{Z[0]}\frac{\int [d\phi] i\psi^{\rho_1}_{\beta_1}(-i\bar{\psi}^{\sigma_1}_{\alpha_1})e^{iS[\phi]}}{Z[0]} \right\}\nonumber\\
&=&\frac{-\delta_{\alpha\beta}}{N_c}\left\{\langle 0|T[\psi^\rho_\beta\bar{\psi}^\sigma_\alpha \psi^{\rho_1}_{\beta_1}\bar{\psi}^{\sigma_1}_{\alpha_1}]|0\rangle_f -
\langle 0|T[\psi^\rho_\beta\bar{\psi}^\sigma_\alpha]|0\rangle_f \langle 0|T[ \psi^{\rho_1}_{\beta_1}\bar{\psi}^{\sigma_1}_{\alpha_1}]|0\rangle_f \right\}\label{delPhi1}
\end{eqnarray}
Since it is closely related to 4-point quark Green's function, it may satisfy an inhomogeneous Bethe-Salpter equation, and we found this is indeed the case.
Taking derivatives on Eq. (\ref{Pidef}) with respect to $\bar{I}I$, we obtain:
\begin{eqnarray}\label{PivsPhi2}
0&=&N_c\bigg[i[i\partial\!\!\!/-M-\Pi_c]^{-1,\rho,\sigma'}\delta\Pi^{\sigma\rho}_{c\rho_1\sigma_1,\beta_1\alpha_1}[i\partial\!\!\!/-M-\Pi_c]^{-1,\rho',\sigma}+\delta\Phi^{\sigma\rho}_{c\rho_1\sigma_1,\beta_1\alpha_1}
\bigg]\nonumber\\
&&+\delta_{\alpha_1\beta_1}[i\partial\!\!\!/-M-\Pi_c]^{-1,\rho,\sigma_1}[i\partial\!\!\!/-M-\Pi_c]^{-1,\rho_1,\sigma}.
\end{eqnarray}
Similarly, from Eq. (\ref{Phidef}), we have
\begin{eqnarray}\label{PhivsPi1}
\delta\Pi^{\sigma\rho}_{c\rho_1\sigma_1\beta_1\alpha_1}&=&-\sum^{\infty}_{n=2}\frac{(-i)^{n}(N_c g^2)^{n-1}}{(n-2)!}\bar{G}^{\sigma\sigma_2\sigma'_3\cdots\sigma'_n}_{\rho\rho_2\rho'_3\cdots\rho'_n}\delta\Phi^{\sigma_2\rho_2}_{c\rho_1\sigma_1\beta_1\alpha_1}
\Phi_c^{\sigma'_3\rho'_3}\cdots\Phi_c^{\sigma'_n\rho'_n}.
\end{eqnarray}
Inserting Eq. (\ref{PhivsPi1}) into Eq. (\ref{PivsPhi2}), we immediately arrive at an inhomogeneous Bethe-Salpeter equation:
\begin{eqnarray}\label{InhomBS}
\delta\Phi^{\sigma\rho}_{c\rho_1\sigma_1,\beta_1\alpha_1}&=&i[i\partial\!\!\!/-M-\Pi_c]^{-1,\rho,\sigma'}\sum^{\infty}_{n=2}\frac{(-i)^{n}(N_c g^2)^{n-1}}{(n-2)!}\bar{G}^{\sigma'\sigma_2\sigma'_3\cdots\sigma'_n}_{\rho'\rho_2\rho'_3\cdots\rho'_n}\delta\Phi^{\sigma_2\rho_2}_{c\rho_1\sigma_1\beta_1\alpha_1}
\Phi_c^{\sigma'_3\rho'_3}\cdots\Phi_c^{\sigma'_n\rho'_n}[i\partial\!\!\!/-M-\Pi_c]^{-1,\rho',\sigma}\nonumber\\
&&-\frac{\delta_{\alpha_1\beta_1}}{N_c}[i\partial\!\!\!/-M-\Pi_c]^{-1,\rho,\sigma_1}[i\partial\!\!\!/-M-\Pi_c]^{-1,\rho_1,\sigma}
\end{eqnarray}

\begin{figure}[hbt]
\centering
\includegraphics[width = 0.8\textwidth]{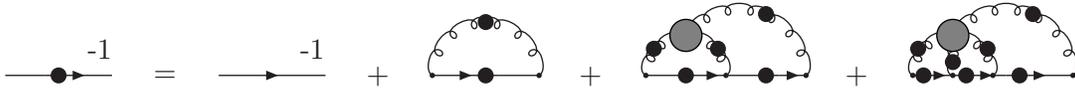}
\caption{The gap equation up to the NNL-order truncation. The black circles indicate the propagators are fully dressed. The gray circles are connected Green's functions. The dots are bare vertices.}\label{gap}
\end{figure}

\begin{figure}[hbt]
\centering
\includegraphics[width = 0.8\textwidth]{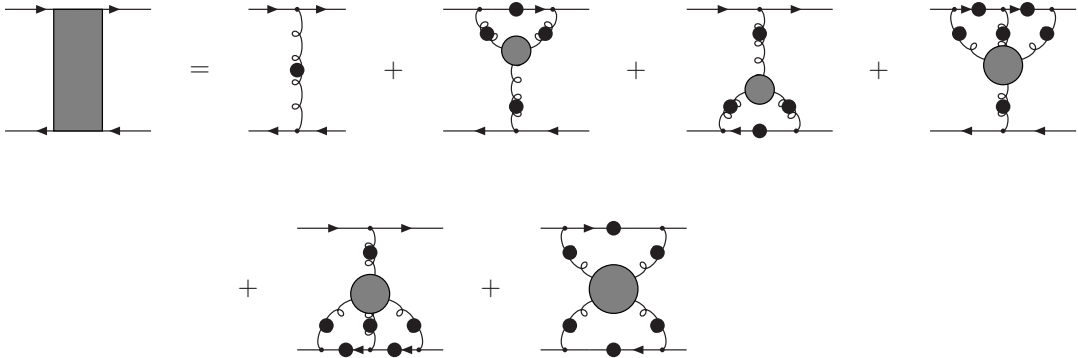}
\caption{The meson BS kernel up to the NNL-order truncation. The gray rectangle is the BS kernel. The black circles indicate the propagators are fully dressed. The gray circles are connected Green's functions. The dots are bare vertices.}\label{BSE}
\end{figure}

Now we can use the standard technique to extract the pole contribution from $\delta\Phi^{\sigma\rho}_{c\rho_1\sigma_1,\beta_1\alpha_1}$, and obtain the homogenous Bethe-Salpeter equation for mesons:
\begin{eqnarray}\label{homBS}
\chi^{\rho\sigma}_{P,s}&=&i[i\partial\!\!\!/-M-\Pi_c]^{-1,\rho,\sigma'}\sum^{\infty}_{n=2}\frac{(-i)^{n}(N_c g^2)^{n-1}}{(n-2)!}\bar{G}^{\sigma'\sigma_2\sigma'_3\cdots\sigma'_n}_{\rho'\rho_2\rho'_3\cdots\rho'_n}
\Phi_c^{\sigma'_3\rho'_3}\cdots\Phi_c^{\sigma'_n\rho'_n}[i\partial\!\!\!/-M-\Pi_c]^{-1,\rho',\sigma}\chi^{\rho_2\sigma_2}_{P,s}\nonumber\\
&=&-i[i\partial\!\!\!/-M-\Pi_c]^{-1,\rho,\sigma'}K^{\sigma'\rho'}_{\sigma_2\rho_2}[i\partial\!\!\!/-M-\Pi_c]^{-1,\rho',\sigma}\chi^{\rho_2\sigma_2}_{P,s}
\end{eqnarray}
where the BS kernel $K^{\sigma'\rho'}_{\sigma_2\rho_2}$ is defined as
\begin{equation}\label{Kernel}
K^{\sigma'\rho'}_{\sigma_2\rho_2}\equiv-\sum^{\infty}_{n=2}\frac{(-i)^{n}(N_c g^2)^{n-1}}{(n-2)!}\bar{G}^{\sigma'\sigma_2\sigma'_3\cdots\sigma'_n}_{\rho'\rho_2\rho'_3\cdots\rho'_n}
\Phi_c^{\sigma'_3\rho'_3}\cdots\Phi_c^{\sigma'_n\rho'_n}=\frac{\delta\Pi_c^{\sigma'\rho'}}{\delta\Phi_c^{\sigma_2\rho_2}}
\end{equation}
From Eq. (\ref{Kernel}), we see that the BS kernel can be obtained by breaking each of all the quark propagators in the quark self energy. Once a truncation being made in the gap equation, the BS kernel should be truncated accordingly. This relation is also shown in studies under symmetry preserving truncations~\cite{BDTR,BCPQR}, so our method exemplifies the chiral symmetry preserving truncation scheme. The lowest order in this truncation scheme is the RL truncation. To beyond the RL truncation, one can include the 3-gluon Green's function contribution, which gives quark-gluon vertex corrections in the gap equation and in the BSE. When the connected 4-gluon Green's functions contributions are included, the gap equation still receives QGV corrections, while the BSE receives corrections due to H-shape diagrams which may be important in scalar and axial vector channels. 
The diagrammatic expressions for the gap equation and the BSE truncated up to the next-to-next-to-leading (NNL) order are shown in Fig. \ref{gap} and Fig. \ref{BSE} respectively.

As indicated in Introduction, RL truncation gives poor results in some channels, such as the scalar and axial-vector channels. The truncation scheme presented here allows us to make improvements in these channels. A few works have studied the impacts of going beyond the RL truncation by taking into account the three-gluon self-interaction ~\cite{FW,WF,AW,FUW}, which corresponds to the next-to-leading-order corrections in our scheme. It was found that at this order, considerable corrections already appear in the scalar and axial-vector channels, and improvements were found in the axial-vector channel~\cite{FW,WF,AW}. 

Our framework automatically provide symmetry preserving truncations. The reason can be understood if we write the meson BSE in another form.
Actually, Eq. (\ref{homBS}) can also be written in terms of derivatives of the effective action $\Gamma[\bar{\Phi},\bar{\Pi}]$ as
\begin{eqnarray}\label{homBS3}
\left\{\bar{\Phi}^{-1,\rho_2\sigma'}\frac{\delta^2\Gamma}{\delta\bar{\Pi}^{\sigma'\rho'}\delta\bar{\Phi}^{\sigma_1\rho_1}}\bar{\Phi}^{-1,\rho'\sigma_2}+i \frac{\delta^2\Gamma}{\delta\bar{\Phi}^{\sigma_2\rho_2}\delta\bar{\Phi}^{\sigma_1\rho_1}}\right\} \bigg|_{\bar{\Phi}=\Phi_c,\bar{\Pi}=\Pi_c}\chi^{\rho_2\sigma_2}_{P,s}&=&0,
\end{eqnarray}
where $\Gamma[\bar{\Phi},\bar{\Pi}]$ is just $\frac{-i}{N_c}\ln Z'|_{\mathrm{e.s.v.}}$ with $\Phi_c,\Pi_c$ replaced with arbitrary bilocal functions $\bar{\Phi},\bar{\Pi}$ respectively, i.e.
\begin{eqnarray}
\Gamma[\bar{\Phi},\bar{\Pi}]&=& -i{\rm Tr}\ln[i\partial\!\!\!/-M-\bar{\Pi}]
+\bar{\Phi}^{\sigma\rho}\bar{\Pi}^{\sigma\rho}+ \sum^{\infty}_{n=2}\frac{(-i)^{n}(N_c g^2)^{n-1}}{n!}\bar{G}^{\sigma_1\cdots\sigma_n}_{\rho_1\cdots\rho_n}\bar{\Phi}^{\sigma_1\rho_1}\cdots\bar{\Phi}^{\sigma_n\rho_n}.
\end{eqnarray}
Equation (\ref{homBS3}) is just the analogue of Eq. (3) in Ref. \cite{Munczek}. We have seen that the BSE and the DSEs all can be derived from the same generating functional, so symmetries respected by the generating functional will be automatically respected by the equations derived from it. Especially, truncations can be made for the generating functional $-i\ln Z'$, then all the derived equations will be in consistent with each other and respect symmetries that retained in the truncated generating functional. This is why we automatically have a chiral symmetry preserving truncation scheme.

For the phenomenological studies focusing on meson properties under RL truncation, the gluon propagator is usually treated as an input, i.e. given as a model or by fitting lattice results, etc.. It avoids complications in dealing with too many coupled equations, thus is very useful for practical use. In our approach, the gauge sector and fermion sector of QCD are treated differently at the very beginning, so it retains the merit just mentioned. In this respect, our approach can be viewed as an extension from considering only the gluon propagator's effects to considering higher-order gluon Green's functions' effects on the fermion sector. It results in truncations beyond the RL approximation on one hand and completes the gauge sector because of the inclusion of the gluon self-interactions on the other.

All the higher-order terms in our truncation scheme are originated from non-Abelian type dynamics, i.e., gluon self-interactions and gluon-ghost interactions. Remember that we have taken large $N_c$ limit, which means that non-Abelian type contributions are always dominant compared to Abelian type contributions up to any order (beyond RL truncation).

\section{The Slavnov-Taylor identity for the quark-gluon vertex}

The Slavnov-Taylor identity for the QGV provides an important constraint among the quark propagator, the QGV and the quark-ghost scattering kernel, which reflects the BRS symmetry of QCD. In principle, a proper truncation should guarantee solutions of truncated DSEs satisfy STIs. However it is not easy to maintain this requirement in practice. In order to see the explicit form of the STI of QGV in our formulism, we derive the quark-ghost scattering kernel in this section. Now, we need to express the Fadeev-Popov determinant in terms of ghost fields $\bar{\phi}_i(x)$ and $\phi_i(x)$, then we have a ghost term in QCD Lagrangian:
\begin{equation}\label{ghostL}
{\mathcal{L}}_{\mathrm{ghost}}=-(\partial_\mu \bar{\phi}^i)D^\mu_{ij}\phi^j
\end{equation}
Using the BRS symmetry of the theory, one can arrive at a STI relating the quark-gluon 3-point Green's function to the quark-antiquark-ghost-antighost 4-pont Green's function~\cite{PT}.
\begin{eqnarray}\label{STI0}
0&=&\omega\bigg\{\frac{i}{2}g \lambda^k_{\alpha\gamma}\langle0|T[\phi_k(x)\psi_\gamma(x)\bar{\psi}_\beta(y)\bar{\phi}_i(z) ]|0\rangle_f -\frac{i}{2}g \lambda^k_{\gamma\beta}\langle0|T[\psi_\alpha(x)\bar{\psi}_\gamma(y)\phi_k(y)\bar{\phi}_i(z) ]|0\rangle_f \nonumber\\
&&+\frac{1}{\xi} \langle0|T[\psi_\alpha(x)\bar{\psi}_\beta(y)\partial_\mu A^\mu_i(z) ]|0\rangle_f \bigg\}.
\end{eqnarray}
We shall see that this is just the STI for QGV.
Introducing the quark-ghost scattering kernels $H_{\alpha\beta}^i$ and $\bar{H}_{\alpha\beta}^i$ according to
\begin{eqnarray}
&&H_{\alpha\gamma}^{k'}(p,q,r) S_{\gamma\beta}(\slashed r) \tilde{D}_{k'i}(q^2)\nonumber\\
&\equiv&
\frac{1}{2} \lambda^k_{\alpha\gamma}\int d^4 x d^4y e^{-i p\cdot x}e^{-i r\cdot y}\langle0|T[\phi_k(x)\psi_\gamma(x)\bar{\psi}_\beta(y)\bar{\phi}_i(0) ]|0\rangle_f
\end{eqnarray}
and
\begin{eqnarray}
&&S_{\alpha\gamma}(-\slashed p) \bar{H}_{\gamma\beta}^{k'}(-r,-q,-p) \tilde{D}_{k'i}(q^2)\nonumber\\
&\equiv&
\frac{1}{2} \lambda^k_{\gamma\beta}\int d^4 x d^4y e^{-i p\cdot x}e^{-i r\cdot y}\langle0|T[\psi_\alpha(x)\bar{\psi}_\gamma(y)\phi_k(y)\bar{\phi}_i(0) ]|0\rangle_f
\end{eqnarray}
where $q=-p-r$ and $\tilde{D}_{ij}(q^2)$ is the ghost propagator.
With these definitions Eq. (\ref{STI0}) becomes (after taking Fourier transformation)
\begin{eqnarray}
&&gH_{\alpha\gamma}^{k'}(p,q,r) S_{\gamma\beta}(\slashed r) \tilde{D}_{k'i}(q^2)-gS_{\alpha\gamma}(-\slashed p) \bar{H}_{\gamma\beta}^{k'}(-r,-q,-p) \tilde{D}_{k'i}(q^2)\nonumber\\
&=&\frac{1}{\xi}S_{\alpha\gamma}(-\slashed p) \big[-ig\Gamma_{\nu,\gamma\alpha}^{j}(q,p,r)\big]
S_{\alpha\beta}(\slashed r) G^{\nu\mu}_{ji}(q) q_\mu.
\end{eqnarray}
Since for the gluon propagator $G^{\nu\mu}_{ji}(q)$, the non-transverse part equals to that of a free gluon propagator, the equation can be reduced. After extracting the color structure, we obtain
\begin{eqnarray}\label{STI3}
S^{-1}(-\slashed p)H(p,q,r)- \bar{H}(-r,-q,-p) S^{-1}(-\slashed r)
&=&-\frac{1}{q^2}q^\nu \Gamma_{\nu}(q,p,r)\tilde{D}^{-1}(q^2),
\end{eqnarray}
which is just the usual form of the STI of QGV.
In QED, ghost fields decouple from other fields, i.e., $H=\bar{H}=1$ and $\tilde{D}^{-1}(q^2)=-q^2$, then the corresponding Ward-Takahashi identity in QED follows:
\begin{eqnarray}\label{WTI3}
S^{-1}(-\slashed p)-  S^{-1}(-\slashed r)
&=&q^\nu \Gamma_{\nu}(q,p,r)
\end{eqnarray}

Without taking large $N_c$ limit, the Slavnov-Taylor identities are of course satisfied. Since the BRS symmetry is independent of the $SU(N_c)$ group parameter $N_c$, they should still hold at large $N_c$ limit. So in our formulism, the solutions of the gap equation, the QGV DSE and the quark-ghost scattering kernel (untruncated equations) satisfy Eq. (\ref{STI3}). In order to give an explicit form of the quark-ghost scattering kernel, or equivalently the quark-antiquark-ghost-antighost 4-pont Green's function, we need to introduce external sources for ghost fields. The appearance of these external sources will largely complicate the form of the generating functional. We put these material in Appendix B. After a lengthy work, we obtain the generating functional with ghost external sources (up to $\bar{\eta} \eta$ order) at large $N_c$ limit as shown in Eq. (\ref{ZprimGhost}).

Now the quark-antiquark-ghost-antighost 4-pont Green's function can be derived directly by calculating $\frac{-i\delta^4\ln Z'}{\delta \bar{\eta}^j\delta (-\eta^i) \delta \bar{I}_\beta \delta (-I_\alpha)}$. Eventually, we obtain
\begin{eqnarray}\label{qgk3}
&&\langle 0|T[\phi_j(\tilde{y})\bar{\phi}_i(\tilde{x})\psi_{\beta}^\rho(y)\bar{\psi}_\alpha^\sigma(x)]|0\rangle_c\nonumber\\
&=&\int d^4x' d^4 y'\bigg[-
(i\partial\!\!\!/-M-\Pi_c)^{-1}_{\rho\rho'}(y,y')\frac{\delta^2\Pi_c^{\rho'\sigma'}(y',x')}{\delta \bar{\eta}^j(\tilde{y})\delta(- \eta^i(\tilde{x}))}(i\partial\!\!\!/-M-\Pi_c)^{-1}_{\sigma'\sigma}(x',x)\bigg]\delta_{\alpha\beta}
\\
&&+\int d^4x' d^4 y'\bigg[(i\partial\!\!\!/-M-\Pi_c)^{-1}_{\rho\rho'}(y,y')\big(\Delta^{b\rho'\sigma'}_{\Phi_c}(\tilde{x},\tilde{y};y',x')[\lambda^i\lambda^j]_{\beta\alpha}\nonumber\\
&&\hspace{2cm}+\Delta^{c\rho'\sigma'}_{\Phi_c}(\tilde{x},\tilde{y};y',x')[\lambda^j\lambda^i]_{\beta\alpha}\big)(i\partial\!\!\!/-M-\Pi_c)^{-1}_{\sigma'\sigma}(x',x)\bigg].\nonumber
\end{eqnarray}
The full Green's function is
\begin{eqnarray}\label{qgk4}
&&\langle 0|T[\phi_j(\tilde{y})\bar{\phi}_i(\tilde{x})\psi_{\beta}^\rho(y)\bar{\psi}_\alpha^\sigma(x)]|0\rangle_f\nonumber\\
&=&i\langle 0|T[\phi_j(\tilde{y})\bar{\phi}_i(\tilde{x})\psi_{\beta}^\rho(y)\bar{\psi}_\alpha^\sigma(x)]|0\rangle_c-\langle 0|T[\psi_{\beta}^\rho(y)\bar{\psi}_\alpha^\sigma(x)]|0\rangle_c\langle 0|T[\phi_j(\tilde{y})\bar{\phi}_i(\tilde{x})]|0\rangle_c\nonumber\\
&=&\int d^4x' d^4 y'\bigg[-
(i\partial\!\!\!/-M-\Pi_c)^{-1}_{\rho\rho'}(y,y')\frac{i\delta^2\Pi_c^{\rho'\sigma'}(y',x')}{\delta \bar{\eta}^j(\tilde{y})\delta(- \eta^i(\tilde{x}))}(i\partial\!\!\!/-M-\Pi_c)^{-1}_{\sigma'\sigma}(x',x)\bigg]\delta_{\alpha\beta}
\nonumber\\
&&+\int d^4x' d^4 y'\bigg[(i\partial\!\!\!/-M-\Pi_c)^{-1}_{\rho\rho'}(y,y')\big(i\Delta^{b\rho'\sigma'}_{\Phi_c}(\tilde{x},\tilde{y};y',x')[\lambda^i\lambda^j]_{\beta\alpha}\nonumber\\
&&\hspace{2cm}+i\Delta^{c\rho'\sigma'}_{\Phi_c}(\tilde{x},\tilde{y};y',x')[\lambda^j\lambda^i]_{\beta\alpha}\big)(i\partial\!\!\!/-M-\Pi_c)^{-1}_{\sigma'\sigma}(x',x)\bigg]\nonumber\\
&&-\delta_{\alpha\beta}\delta_{ij}(i\partial\!\!\!/-M-\Pi_c)^{-1}_{\rho\sigma}(y,x)\tilde{D}(\tilde{x},\tilde{y}).
\end{eqnarray}
Taking Eq. (\ref{QuarkGluonVertex-1}) and Eq. (\ref{qgk4}) into Eq. (\ref{STI0}), we obtain
\begin{eqnarray}\label{STI-final}
&&\lambda^i_{\alpha\beta}(\frac{ig}{2})\int d^4x' d^4 y'(i\partial\!\!\!/-M-\Pi_c)^{-1}_{\rho\rho'}(x,y')\bigg\{-\frac{i\delta^2\Pi_c^{\rho'\sigma'}(y',x')}{\delta \bar{\eta}^j(x)\delta(- \eta^i(z))}+\frac{i\delta^2\Pi_c^{\rho'\sigma'}(y',x')}{\delta \bar{\eta}^j(y)\delta(- \eta^i(z))}\nonumber\\
&&\hspace*{6cm}+2N_c\big(i\Delta^{c\rho'\sigma'}_{\Phi_c}(z,x;y',x')
-i\Delta^{b\rho'\sigma'}_{\Phi_c}(z,y;y',x')\big)\nonumber\\
&&\hspace*{6cm}+(i\partial\!\!\!/-M-\Pi_c)_{\rho'\sigma'}(y',x')\big(\tilde{D}(z,x)-\tilde{D}(z,y)\big)\bigg\}(i\partial\!\!\!/-M-\Pi_c)^{-1}_{\sigma'\sigma}(x',y)\nonumber\\
&=&\lambda^i_{\alpha\beta}\int d^4x' d^4 y'(i\partial\!\!\!/-M-\Pi_c)^{-1}_{\rho\rho'}(x,y')\bigg[-\frac{1}{\xi}\frac{\partial}{\partial z^\mu}\Delta_{\Phi_c}^{\mu\rho'\sigma'}(z,y',x')\bigg](i\partial\!\!\!/-M-\Pi_c)^{-1}_{\sigma'\sigma}(x',y).
\end{eqnarray}
This is the STI for QGV in our formalism.

The undetermined function $\frac{\delta^2\Pi_c^{\rho'\sigma'}(y',x')}{\delta \bar{\eta}^j(x)\delta(- \eta^i(z))}$ satisfies coupled integral equations, which, in a compact form, reads
\begin{eqnarray}
\frac{\delta^2\Pi_c}{\delta \bar{\eta}^j\delta(- \eta^i)}&=&i\frac{\delta_{ij}}{2N_c}\mathrm{Tr}\big[(i\partial\!\!\!/-M-\Pi_c)^{-1}\big(\frac{\delta\Delta^{b}_{\Phi_c}}{\delta\Phi_c}
+\frac{\delta\Delta^{c}_{\Phi_c}}{\delta\Phi_c}\big)\big]\nonumber\\
&&\hspace*{-1cm}-\frac{\delta_{ij}}{N_c}\frac{\delta\Delta^a_{\Phi_c}}{\delta\Phi_c}-\frac{\delta \Delta^{d}_{\Phi_c}}{N_c\delta\Phi_c}\langle\bar{\psi}\lambda^i\psi\rangle\langle\bar{\psi}\lambda^j\psi\rangle\nonumber\\
&&\hspace*{-1cm}+\sum^{\infty}_{n=2}\frac{(-i)^n (N_cg^2)^{n-1}}{(n-2)!}\bar{G}_{n-1}\frac{\delta^2\Phi_c}{\delta \bar{\eta}^j\delta(- \eta^i)}
\Phi_c\cdots\Phi_c,
\end{eqnarray}
\begin{eqnarray}
\frac{\delta^2\Phi_c}{\delta \bar{\eta}^j\delta(- \eta^i)}&=&-i(i\partial\!\!\!/-M-\Pi_c)^{-1}\frac{\delta^2\Pi_c}{\delta \bar{\eta}^j\delta(- \eta^i)}(i\partial\!\!\!/-M-\Pi_c)^{-1}\nonumber\\
&&\hspace*{-1cm}+i\frac{\delta_{ij}}{2N_c}\mathrm{Tr}\big[(i\partial\!\!\!/-M-\Pi_c)^{-1}\big(\Delta^{b}_{\Phi_c}
+\Delta^{c}_{\Phi_c}\big)\big].
\end{eqnarray}
$\Delta_{\Phi_c}^a,\Delta_{\Phi_c}^b,\Delta_{\Phi_c}^c$ and $\Delta_{\Phi_c}^d$ in the l.h.s. of Eq. (\ref{STI-final}) are integrations of 2-ghost+n-gluon Green's functions; while $\Delta_{\Phi_c}^{\mu}$ in the r.h.s. of Eq. (\ref{STI-final}) is an integration of n-gluon Green's functions. These Green's functions are not independent. Actually, they satisfy their own DSEs and STIs derived from the pure Yang-Mills part of the theory.

We have given the expression of the quark-antiquark-ghost-antighost 4-pont Green's function (Eq. (\ref{qgk4})), with which the quark-ghost scattering kernel is defined, and the STI for the quark-gluon vertex in coordination space (Eq. (\ref{STI-final})) in our formulism. As discussed before, the solutions of untruncated DSEs at large $N_c$ limit should satisfy Eq. (\ref{STI-final}). More interesting case arise when a truncation is made in the DSEs. The simplest truncation is the Abelian approximation (rainbow approximation). In this case, the ghost fields can decouple from other fields, so the QGV STI reduces to WTI, and one can check that the gauge-fermion vertex WTI is not satisfied under rainbow approximation. In principle, the STI for the QGV at higher-order truncations can be tested using Eq. (\ref{STI-final}). Unfortunately, we can not do it due to the complicated forms of relevant Green's functions. However, we hope the discussion made in this section can reveal some hints in the studies of QGV's STI.

\section{Summary}
We introduced an alternate form for the QCD generating functional which is a generalization of the one used in a previous study. This form has the power to address various nonperturbative problems in QCD. Specifically, we employed it to study Dyson-Schwinger Equations and Bethe-Salpeter Equation and their truncations. The large $N_c$ expansion was taken and we were concentrated on the large $N_c$ limit. Under large $N_c$ limit, a systematic order-by-order truncation scheme with all higher-order terms explicitly given was proposed. One benefit of this framework is that one can make truncations in the generating functional from which all the equations are derived. So any linearly presented symmetry preserved in the truncated generating functional can automatically transmit to the DSEs and the BSE. To be specific, truncations can be made by keeping finite terms in the expansion of the generating functional with respect to the number of external legs of connected gluon Green's functions. With any such truncation, the chiral symmetry is conserved (in the chiral limit). So the truncated DSEs and BSE preserve chiral symmetry, and the truncation scheme proposed here is actually a symmetry preserving truncation scheme. Another benefit is that our truncation scheme avoids ambiguities appearing in the methods making direct use of QGV DSE.

Since the explicit forms of the DSEs and the BSE are established at large $N_c$ limit, the integration kernels suffer from corrections of $1/N_c$ order. The positive side of taking large $N_c$ limit is that all terms in the integration kernels making corrections to the RL truncation are non-Abelian type, i.e., they have no counterparts in QED, which means that non-Abelian type contributions are always dominant compared to Abelian type contributions up to any order (beyond RL truncation), and it is especially useful for testing non-Abelian dynamics. In this framework, the H-shape diagram in the BSE kernel, which is considered important in the scalar and axial vector channels, will appear when connected 4-gluon Green's functions are present. Meanwhile, the 4-gluon self-interaction appears at the same order, so truncation made up to this order (i.e., keep the gluon propagator term, the 3-gluon self-interaction term and the connected 4-gluon Green's functions term in the kernels) is of great interests.

In order to study the QGV STI, we derived the quark-ghost scattering kernel. Ghost fields are explicitly shown in the generating functional and the external sources are introduced. Although we cannot verify the STI directly due to the complicated form of the quark-ghost scattering kernel, we hope those discussions could shed some light on further studies and on the modeling of QGV using the STI.

%%%%%%%%%%%%%%%%%%%%%%%%%%%%%%%%%%%%%%%%%%%%%%%%%%%%%%%%%%%%%%%
 \appendix
  \renewcommand{\appendixname}{Appendix}

\section{}
In this appendix, we give a detailed discussion on the Fierz reordering. In order to obtain Eq. (\ref{Ext2Gluon}), we note that
\begin{eqnarray}                             %(4)
 &&G_{\mu_1\mu_2}^{i_1i_2}(x_1,x_2)[-g\bar{\psi}^{a_1}_{{\alpha}_1}(x_1) (\frac{\lambda_{i_1}}{2})_{\alpha_1\beta_1}\gamma^{\mu_1}
 {\psi}^{a_1}_{{\beta}_1}(x_1)][-g\bar{\psi}^{a_2}_{{\alpha}_2}(x_2)(\frac{\lambda_{i_2}}{2})_{\alpha_2\beta_2}\gamma^{\mu_2}{\psi}^{a_2}_{\beta_2}(x_2)]\nonumber\\
 &&=G_{\mu_1\mu_2}(x_1,x_2)[-g\bar{\psi}^{a_1}_{{\alpha}_1}(x_1)\gamma^{\mu_1}{\psi}^{a_1}_{{\beta}_1}(x_1)][-g\bar{\psi}^{a_2}_{{\alpha}_2}(x_2)\gamma^{\mu_2}{\psi}^{a_2}_{\beta_2}(x_2)]
 \frac{1}{2}(\delta_{\alpha_1\beta_2}\delta_{\alpha_2\beta_1}-\frac{1}{N_c}\delta_{\alpha_1\beta_1}\delta_{\alpha_2\beta_2})\nonumber\\
 &&=\int d^4x_1'd^4x_2'~g^2G_{\mu_1\mu_2}(x_1,x_2)[-\frac{1}{2}(\gamma^{\mu_1})_{\sigma_1\rho_2}(\gamma^{\mu_2})_{\sigma_2\rho_1}\delta(x_1'-x_2)\delta(x_2'-x_1)\nonumber\\
 &&\hspace*{0.5cm}-\frac{1}{2N_c}(\gamma^{\mu_1})_{\sigma_1\rho_1}(\gamma^{\mu_2})_{\sigma_2\rho_2}\delta(x_1'-x_1)\delta(x_2'-x_2)]
 \bar{\psi}_{\alpha_1}^{\sigma_1}(x_1)\psi_{\alpha_1}^{\rho_1}(x_1') \bar{\psi}_{\alpha_2}^{\sigma_2}(x_2)\psi_{\alpha_2}^{\rho_2}(x_2').
\end{eqnarray}
In obtaining the above result, we have used relations
\begin{eqnarray}
 &&G_{\mu_1\mu_2}^{i_1i_2}(x_1,x_2)=\delta^{i_1i_2}G_{\mu_1\mu_2}(x_1,x_2),\\
 &&(\frac{\lambda_i}{2})_{\alpha_1\beta_1}(\frac{\lambda_i}{2})_{\alpha_2\beta_2}= \frac{1}{2}(\delta_{\alpha_1\beta_2}\delta_{\alpha_2\beta_1}-\frac{1}{N_c}\delta_{\alpha_1\beta_1}\delta_{\alpha_2\beta_2}).\label{GellMannMatrix}
\end{eqnarray}
Hence the extended 2-point Green's function is
\begin{eqnarray}
\overline{G}_{\rho_1\rho_2}^{\sigma_1\sigma_2}(x_1,x_1',x_2,x_2')=-\frac{1}{2}G_{\mu_1\mu_2}(x_1,x_2)[&&(\gamma^{\mu_1})_{\sigma_1\rho_2}(\gamma^{\mu_2})_{\sigma_2\rho_1}
\delta(x_1'-x_2)\delta(x_2'-x_1)+\nonumber\\
&&+\frac{1}{N_c}(\gamma^{\mu_1})_{\sigma_1\rho_1}(\gamma^{\mu_2})_{\sigma_2\rho_2}\delta(x_1'-x_1)\delta(x_2'-x_2)].
\end{eqnarray}
For the 3-point functions, we need the $SU(N_c)$ relations
\begin{eqnarray}
f_{ijk}&=&-2i([\frac{\lambda_i}{2},\frac{\lambda_j}{2}]\frac{\lambda_k}{2})=\frac{i}{4}(\lambda_j)_{\alpha\beta}(\lambda_i)_{\beta\gamma}(\lambda_k)_{\gamma\alpha}-
\frac{i}{4}(\lambda_i)_{\alpha\beta}(\lambda_j)_{\beta\gamma}(\lambda_k)_{\gamma\alpha}\nonumber\\
d_{ijk}&=&-2i(\{\frac{\lambda_i}{2},\frac{\lambda_j}{2}\}\frac{\lambda_k}{2})=\frac{1}{4}(\lambda_j)_{\alpha\beta}(\lambda_i)_{\beta\gamma}(\lambda_k)_{\gamma\alpha}+
\frac{1}{4}(\lambda_i)_{\alpha\beta}(\lambda_j)_{\beta\gamma}(\lambda_k)_{\gamma\alpha}
\end{eqnarray}
then
\begin{eqnarray}
&&f_{ijk}(\lambda_j)_{\alpha_2\beta_2}(\lambda_k)_{\alpha_3\beta_3}=i[\delta_{\alpha_3\beta_2}(\lambda_i)_{\alpha_2\beta_3}-\delta_{\alpha_2\beta_3}(\lambda_i)_{\alpha_3\beta_2}]\\
&&d_{ijk}(\lambda_j)_{\alpha_2\beta_2}(\lambda_k)_{\alpha_3\beta_3}=\delta_{\alpha_3\beta_2}(\lambda_i)_{\alpha_2\beta_3}+\delta_{\alpha_2\beta_3}(\lambda_i)_{\alpha_3\beta_2}
-\frac{2}{N_c}\delta_{\alpha_3\beta_3}(\lambda_i)_{\alpha_2\beta_2}-\frac{2}{N_c}\delta_{\alpha_2\beta_2}(\lambda_i)_{\alpha_3\beta_3}
\end{eqnarray}
and
\begin{eqnarray}
&&f_{ijk}(\lambda_i)_{\alpha_1\beta_1}(\lambda_j)_{\alpha_2\beta_2}(\lambda_k)_{\alpha_3\beta_3}=2i(\delta_{\alpha_3\beta_2}\delta_{\alpha_2\beta_1}\delta_{\alpha_1\beta_3}-
\delta_{\alpha_3\beta_1}\delta_{\alpha_1\beta_2}\delta_{\alpha_2\beta_3})\\
&&d_{ijk}(\lambda_i)_{\alpha_1\beta_1}(\lambda_j)_{\alpha_2\beta_2}(\lambda_k)_{\alpha_3\beta_3}=2(\delta_{\alpha_3\beta_2}\delta_{\alpha_2\beta_1}\delta_{\alpha_1\beta_3}+
\delta_{\alpha_3\beta_1}\delta_{\alpha_1\beta_2}\delta_{\alpha_2\beta_3})\nonumber\\
&&\hspace*{4cm}-\frac{2}{N_c}\delta_{\alpha_1\beta_1}\delta_{\alpha_2\beta_3}\delta_{\alpha_3\beta_2}
-\frac{2}{N_c}\delta_{\alpha_1\beta_3}\delta_{\alpha_2\beta_2}\delta_{\alpha_3\beta_1}-\frac{2}{N_c}\delta_{\alpha_1\beta_2}\delta_{\alpha_2\beta_1}\delta_{\alpha_3\beta_3}
+\frac{4}{N_c^2}\delta_{\alpha_1\beta_1}\delta_{\alpha_2\beta_2}\delta_{\alpha_3\beta_3}~~~
\end{eqnarray}
The 3-point Green's function can be written as
\begin{eqnarray}
G^{i_1i_2i_3}_{\mu_1\mu_2\mu_3}(x_1,x_2,x_3)=gf_{i_1i_2i_3}G^{(0)}_{\mu_1\mu_2\mu_3}(x_1,x_2,x_3)+gd_{i_1i_2i_3}G^{(1)}_{\mu_1\mu_2\mu_3}(x_1,x_2,x_3)
\end{eqnarray}
Combine them together, we get
\begin{eqnarray}
&&\overline{G}^{\sigma_1\sigma_2\sigma_3}_{\rho_1\rho_2\rho_3}(x_1,x_1',x_2,x_2',x_3,x_3')\nonumber\\
&&=\frac{i}{4}G^{(0)}_{\mu_1\mu_2\mu_3}(x_1,x_2,x_3)[\gamma^{\mu_1}_{\sigma_1\rho_2}\gamma^{\mu_2}_{\sigma_2\rho_3}\gamma^{\mu_3}_{\sigma_3\rho_1}\delta(x_1'-x_3)
\delta(x_2'-x_1)\delta(x_3'-x_2)
-\gamma^{\mu_1}_{\sigma_1\rho_3}\gamma^{\mu_2}_{\sigma_2\rho_1}\gamma^{\mu_3}_{\sigma_3\rho_2}\delta(x_1'-x_2)
\delta(x_2'-x_3)\delta(x_3'-x_1)]\nonumber\\
&&\hspace*{0.5cm}+\frac{1}{4}G^{(1)}_{\mu_1\mu_2\mu_3}(x_1,x_2,x_3)[\gamma^{\mu_1}_{\sigma_1\rho_3}\gamma^{\mu_2}_{\sigma_2\rho_1}\gamma^{\mu_3}_{\sigma_3\rho_2}\delta(x_1'-x_2)
\delta(x_2'-x_3)\delta(x_3'-x_1)
+\gamma^{\mu_1}_{\sigma_1\rho_2}\gamma^{\mu_2}_{\sigma_2\rho_3}\gamma^{\mu_3}_{\sigma_3\rho_1}\delta(x_1'-x_3)
\delta(x_2'-x_1)\delta(x_3'-x_2)\nonumber\\
&&\hspace*{0.5cm}+\frac{2}{N_c}\gamma^{\mu_1}_{\sigma_1\rho_1}\gamma^{\mu_2}_{\sigma_2\rho_3}\gamma^{\mu_3}_{\sigma_3\rho_2}\delta(x_1'-x_1)\delta(x_2'-x_3)\delta(x_3'-x_2)
+\frac{2}{N_c}\gamma^{\mu_1}_{\sigma_1\rho_3}\gamma^{\mu_2}_{\sigma_2\rho_2}\gamma^{\mu_3}_{\sigma_3\rho_1}\delta(x_1'-x_3)\delta(x_2'-x_2)\delta(x_3'-x_1)\nonumber\\
&&\hspace*{0.5cm}+\frac{2}{N_c}\gamma^{\mu_1}_{\sigma_1\rho_2}\gamma^{\mu_2}_{\sigma_2\rho_1}\gamma^{\mu_3}_{\sigma_3\rho_3}\delta(x_1'-x_2)\delta(x_2'-x_1)\delta(x_3'-x_3)
+\frac{4}{N_c^2}\gamma^{\mu_1}_{\sigma_1\rho_1}\gamma^{\mu_2}_{\sigma_2\rho_2}\gamma^{\mu_3}_{\sigma_3\rho_3}\delta(x_1'-x_1)\delta(x_2'-x_2)\delta(x_3'-x_3)]
\end{eqnarray}

In general, $\overline{G}^{\sigma_1\ldots\sigma_n}_{\rho_1\ldots\rho_n}(x_1,x_1',\ldots,x_n,x_n')$ has following properties:
\begin{itemize}
\item the basic term is $\gamma^{\mu_1}_{\sigma_1\rho_{i_1}}\delta(x_1-x'_{i_1})\cdots\gamma^{\mu_n}_{\sigma_n\rho_{i_n}}\delta(x_n-x'_{i_n})$ with some nonlocal coefficient which depend on indices $\mu_1,\dots,\mu_n$ and space-time coordinates $x_1,\ldots,x_n$.
\item different terms are corresponding to different arrangement of number $1,\dots,n$ to $i_1,\ldots,i_n$.
\item there is no $\gamma^\mu_{ii}\delta(x_i-x'_i)$ factor.
\end{itemize}

For the extended Green's functions $\tilde{G}_\mu$'s, the 2-point and 3-point Green's functions are, respectively,
\begin{eqnarray}
\tilde{G}^{\sigma}_{\mu_1,\rho}(x_1,x_1',x_2,x'_2)=\frac{1}{2}\gamma^{\mu_2}_{\sigma\rho}G_{\mu_1\mu_2}(x_1,x_2)\delta(x_1'-x_2)\delta(x_2'-x_2)
\end{eqnarray}
and
\begin{eqnarray}
&&\tilde{G}^{\sigma\sigma_3}_{\mu_1,\rho\rho_3}(x_1,x_1',x_2,x'_2,x_3,x'_3)\nonumber\\
&&=\frac{1}{4}\bigg[iG^{(0)}_{\mu_1\mu_2\mu_3}(x_1,x_2,x_3)
\bigg(-\gamma^{\mu_2}_{\sigma\rho_3}\gamma^{\mu_3}_{\sigma_3\rho}\delta(x_1'-x_2)\delta(x_2'-x_3)\delta(x_3'-x_2)
 +\gamma^{\mu_2}_{\sigma_3\rho}\gamma^{\mu_3}_{\sigma\rho_3}\delta(x_1'-x_3)\delta(x_2'-x_2)\delta(x_3'-x_3)]\bigg)\nonumber\\
&&\hspace*{0.3cm} +G^{(1)}_{\mu_1\mu_2\mu_3}(x_1,x_2,x_3)\bigg(-\gamma^{\mu_2}_{\sigma\rho_3}\gamma^{\mu_3}_{\sigma_3\rho}\delta(x_1'-x_2)\delta(x_2'-x_3)\delta(x_3'-x_2)
-\gamma^{\mu_2}_{\sigma_3\rho}\gamma^{\mu_3}_{\sigma\rho_3}\delta(x_1'-x_3)\delta(x_2'-x_2)\delta(x_3'-x_3)\nonumber\\
&&\hspace*{0.3cm}-\frac{4}{N_c}\gamma^{\mu_2}_{\sigma\rho}\gamma^{\mu_3}_{\sigma_3\rho_3}\delta(x_1'-x_2)\delta(x_2'-x_2)\delta(x_3'-x_3)\bigg)\bigg].
\end{eqnarray}

\section{}
We rewrite the original generating functional with ghost fields written explicitly and introducing external sources for them:
\begin{eqnarray}
&&Z[J,{\cal I},\bar{I},I,\bar{\eta}^i,\eta^i]\nonumber\\
&=&\int{\cal D}\psi{\cal D}\bar{\psi}\exp\bigg\{i\int d^4 x
\{\bar{\psi}(i\partial\!\!\!/ +J)\psi+\bar{I}\psi+\bar{\psi}I\}\bigg\}\nonumber\\
&&\times\int{\cal D}A_{\mu}{\cal D}\phi {\cal D}\bar{\phi}\exp\bigg\{ i{\int}d^{4}x\bigg[{\cal L}_{G}(A)
-\frac{1}{2\xi}[F^i(A_{\mu})]^{2}-(\partial_\mu \bar{\phi}^i)D^\mu_{ij}\phi^j+{\cal I}_i^{\prime\mu}A^i_{\mu}+\bar{\eta}^i\phi^i+\bar{\phi}^i \eta^i\bigg]\bigg\}. \label{QCDghost1}
\end{eqnarray}
Formally integrating out the gauge fields and ghost fields, we obtain
\begin{eqnarray}
 &&\int{\cal D}A_{\mu}{\cal D}\phi {\cal D}\bar{\phi}\exp\bigg\{ i{\int}d^{4}x\bigg[{\cal L}_{G}(A)
-\frac{1}{2\xi}[F^i(A_{\mu})]^{2}-(\partial_\mu \bar{\phi}^i)D^\mu_{ij}\phi^j+{\cal I}_i^{\prime\mu}A^i_{\mu}+\bar{\eta}^i\phi^i+\bar{\phi}^i \eta^i\bigg]\bigg\} \nonumber\\
 &&=\exp\;i\sum^{\infty}_{n=2}{\int}d^{4}x_1\cdots{d^{4}x_n}
\frac{i^n}{n!} G_{\mu_1\cdots\mu_n}^{i_1\cdots i_n}(x_1,\cdots,x_n)[\eta^i,\bar{\eta}^j]{\cal I}^{\prime\mu_1}_{i_1}(x_1)\cdots{{\cal I}^{\prime\mu_{n}}_{i_n}(x_n)},\label{Newstart}
\end{eqnarray}
where $G_{\mu_1\cdots\mu_n}^{i_1\cdots i_n}(x_1,\cdots,x_n)[\eta^i,\bar{\eta}^j]$ is connected n-point gluon Green's function depending on $\eta^i,\bar{\eta}^j$. Then the color structure of these Green's functions involve $\eta^i,\bar{\eta}^j$. For example the 2-point Green's function $G^{ij}_{\mu\nu}$ is no longer proportional to $\delta^{ij}$, it also includes terms proportional to $\eta^i\bar{\eta}^j$.

Note that the derivation of STI for quark-gluon vertex only need bilinear term of ghost sources, therefore we do not need to give general expression of (\ref{Newstart}), instead we only keep those terms with ghost sources at most bilinear. In this simplified case, the color structure of $ G_{\mu_1\cdots\mu_n}^{i_1\cdots i_n}(x_1,\cdots,x_n)[\eta^i,\bar{\eta}^j]$ can be explicitly figured out.
\begin{eqnarray}
&&\int{\cal D}A_{\mu}{\cal D}\phi {\cal D}\bar{\phi}\exp\bigg\{ i{\int}d^{4}x\bigg[{\cal L}_{G}(A)
-\frac{1}{2\xi}[F^i(A_{\mu})]^{2}-(\partial_\mu \bar{\phi}^i)D^\mu_{ij}\phi^j+{\cal I}_i^{\prime\mu}A^i_{\mu}+\bar{\eta}^i\phi^i+\bar{\phi}^i \eta^i\bigg]\bigg\} \nonumber\\
&&=\int{\cal D}A_{\mu}\Delta_F(A_\mu)\exp\bigg\{ i{\int}d^{4}x\bigg[{\cal L}_{G}(A)
-\frac{1}{2\xi}[F^i(A_{\mu})]^{2}+{\cal I}_i^{\prime\mu}A^i_{\mu}\bigg]-\bar{\eta}(\partial_{\mu}D^\mu)^{-1}\eta\bigg\}\\
&&=\int{\cal D}A_{\mu}\Delta_F(A_\mu)\exp\bigg[i{\int}d^{4}x\bigg({\cal L}_{G}(A)
-\frac{1}{2\xi}[F^i(A_{\mu})]^{2}+{\cal I}_i^{\prime\mu}A^i_{\mu}\bigg)\bigg]\bigg\{1-\nonumber\\
&&\frac{\int d^4xd^4y\bar{\eta}^i(x)\eta^j(y)
\int{\cal D}A_{\mu}\Delta_F(A_\mu)(\partial_{\mu}D^\mu)^{-1,ij}(x,y)\exp\bigg[i{\int}d^{4}x\bigg({\cal L}_{G}(A)
-\frac{1}{2\xi}[F^i(A_{\mu})]^{2}+{\cal I}_i^{\prime\mu}A^i_{\mu}\bigg)\bigg]}{\int{\cal D}A_{\mu}\Delta_F(A_\mu)\exp\bigg[i{\int}d^{4}x\bigg({\cal L}_{G}(A)
-\frac{1}{2\xi}[F^i(A_{\mu})]^{2}+{\cal I}_i^{\prime\mu}A^i_{\mu}\bigg)\bigg]}+O((\eta\bar{\eta})^2)\bigg\}\nonumber\\
&&=\int{\cal D}A_{\mu}\Delta_F(A_\mu)\exp\bigg[i{\int}d^{4}x\bigg({\cal L}_{G}(A)
-\frac{1}{2\xi}[F^i(A_{\mu})]^{2}+{\cal I}_i^{\prime\mu}A^i_{\mu}\bigg)\bigg]\nonumber\\
&&\times\exp\bigg\{-\frac{\int d^4xd^4y\bar{\eta}^i(x)\eta^j(y)
\int{\cal D}A_{\mu}\Delta_F(A_\mu)(\partial_{\mu}D^\mu)^{-1,ij}(x,y)\exp\bigg[i{\int}d^{4}x\bigg({\cal L}_{G}(A)
-\frac{1}{2\xi}[F^i(A_{\mu})]^{2}+{\cal I}_i^{\prime\mu}A^i_{\mu}\bigg)\bigg]}{\int{\cal D}A_{\mu}\Delta_F(A_\mu)\exp\bigg[i{\int}d^{4}x\bigg({\cal L}_{G}(A)
-\frac{1}{2\xi}[F^i(A_{\mu})]^{2}+{\cal I}_i^{\prime\mu}A^i_{\mu}\bigg)\bigg]}+O((\eta\bar{\eta})^2)\bigg\}.\nonumber
\end{eqnarray}

We define
\begin{eqnarray}
&&-\frac{\int d^4xd^4y\bar{\eta}^i(x)\eta^j(y)
\int{\cal D}A_{\mu}\Delta_F(A_\mu)(\partial_{\mu}D^\mu)^{-1,ij}(x,y)\exp\bigg[i{\int}d^{4}x\bigg({\cal L}_{G}(A)
-\frac{1}{2\xi}[F^i(A_{\mu})]^{2}+{\cal I}_i^{\prime\mu}A^i_{\mu}\bigg)\bigg]}{\int{\cal D}A_{\mu}\Delta_F(A_\mu)\exp\bigg[i{\int}d^{4}x\bigg({\cal L}_{G}(A)
-\frac{1}{2\xi}[F^i(A_{\mu})]^{2}+{\cal I}_i^{\prime\mu}A^i_{\mu}\bigg)\bigg]}\nonumber\\
=&&i\sum^{\infty}_{n=0}{\int}d^4xd^4yd^{4}x_1\cdots{d^{4}x_n}
\frac{i^n}{n!} \mathcal{G}_{\mu_1\cdots\mu_n}^{(n)i,j;i_1\cdots i_n}(x,y;x_1,\cdots,x_n)\bar{\eta}^i(x)\eta^j(y){\cal I}^{\prime\mu_1}_{i_1}(x_1)\cdots{{\cal I}^{\prime\mu_{n}}_{i_n}(x_n)},
\end{eqnarray}
where $\mathcal{G}^{(n)}$ is just the connected ghost-antighost+n-gluon Green's function.
Since ${\cal I}_i^{\prime\mu}\equiv{\cal I}_i^{\mu}-g\bar{\psi}\frac{\lambda_i}{2}{\gamma}^{\mu}{\psi}$, we can expand the n ${\cal I}^{\prime}$'s at each term and obtain terms of ${\cal I}^0$ order, ${\cal I}^1$ order, $\cdots$, ${\cal I}^n$ order. For our purpose, we only need to give explicit expressions of the ${\cal I}^0$-order terms which read
\begin{eqnarray}
i\sum^{\infty}_{n=0}{\int}d^4xd^4yd^{4}x_1\cdots{d^{4}x_n}
\frac{i^n}{n!} \mathcal{G}_{\mu_1\cdots\mu_n}^{(n)i,j;i_1\cdots i_n}(x,y;x_1,\cdots,x_n)\bar{\eta}^i(x)\eta^j(y)(-g\bar{\psi}\frac{\lambda_{i_1}}{2}{\gamma}^{\mu_1}{\psi}(x_1))\cdots(-g\bar{\psi}\frac{\lambda_{i_n}}{2}{\gamma}^{\mu_n}{\psi}(x_n)).
\end{eqnarray}
In order to do Fierz rearrangement, we need to show color factors explicitly.
In general, terms including $\mathcal{G}^{(2)}$ have two free color indices of the adjoint representation $i,j$ and 4 free color indices of the fundamental representation $\alpha,\beta,\delta,\gamma$, so the most general forms of color factors include
\begin{equation}
\delta_{ij}\delta_{\alpha\beta}\delta_{\delta\gamma},~~~~\delta_{\alpha\beta}[\lambda^{i}\lambda^{j}]_{\delta\gamma},~~~~\delta_{\alpha\beta}[\lambda^{j}\lambda^{i}]_{\delta\gamma},~~~~
\lambda^{i}_{\alpha\beta}\lambda^{j}_{\delta\gamma}.
\end{equation}
For higher orders $\mathcal{G}^{(n)}$'s $(n>2)$, the forms of color factors would be different from them by multiplying $\delta_{\alpha_1\beta_1}\cdots$. For example $\mathcal{G}^{(3)}$ should have $\delta_{ij}\delta_{\alpha\beta}\delta_{\delta\gamma}\delta_{\alpha_1\beta_1}$, $\delta_{\alpha\beta}[\lambda^{i}\lambda^{j}]_{\delta\gamma}\delta_{\alpha_1\beta_1}$, $\delta_{\alpha\beta}[\lambda^{j}\lambda^{i}]_{\delta\gamma}\delta_{\alpha_1\beta_1}$ and $\lambda^{i}_{\alpha\beta}\lambda^{j}_{\delta\gamma}\delta_{\alpha_1\beta_1}$. So $\mathcal{G}^{(n)}$ decomposes into
\begin{equation}
\mathcal{G}^{(n)}=\mathcal{G}^{(na)}[\delta_{ij}\delta_{\alpha\beta}\delta_{\delta\gamma}\delta_{\alpha_1\beta_1}\cdots]
+\mathcal{G}^{(nb)}[\delta_{\alpha\beta}[\lambda^{i}\lambda^{j}]_{\delta\gamma}\delta_{\alpha_1\beta_1}\cdots]
+\mathcal{G}^{(nc)}[\delta_{\alpha\beta}[\lambda^{j}\lambda^{i}]_{\delta\gamma}\delta_{\alpha_1\beta_1}\cdots]+\mathcal{G}^{(nd)}
[\lambda^{i}_{\alpha\beta}\lambda^{j}_{\delta\gamma}\delta_{\alpha_1\beta_1}\cdots].
\end{equation}

So we can write
\begin{eqnarray}
&&i\sum^{\infty}_{n=0}{\int}d^4xd^4yd^{4}x_1\cdots{d^{4}x_n}
\frac{i^n}{n!} \mathcal{G}_{\mu_1\cdots\mu_n}^{(n)i,j;i_1\cdots i_n}(x,y;x_1,\cdots,x_n)\bar{\eta}^i(x)\eta^j(y)\nonumber\\
&&\times(-g\bar{\psi}\frac{\lambda_{i_1}}{2}{\gamma}^{\mu_1}{\psi}(x_1))\cdots(-g\bar{\psi}\frac{\lambda_{i_n}}{2}{\gamma}^{\mu_n}{\psi}(x_n))\nonumber\\
=i&&\left\{{\int}d^4xd^4y\Delta^a_\Phi(x,y)\bar{\eta}^i(x)\eta^i(y)+{\int}d^4xd^4yd^{4}z_1d^4z'_1\Delta^{b\sigma\rho}_\Phi(x,y;z_1,z'_1)
\bar{\psi}^{\sigma}(z_1)\tilde{\bar{\eta}}(x)\tilde{\eta}(y){\psi}^{\rho}(z'_1)\right.\nonumber\\
&&-{\int}d^4xd^4yd^{4}z_1d^4z'_1\Delta^{c\sigma\rho}_\Phi(x,y;z_1,z'_1)
\bar{\psi}^{\sigma}(z_1)\tilde{\eta}(y)\tilde{\bar{\eta}}(x){\psi}^{\rho}(z'_1)\nonumber\\
&&\left.+{\int}d^4xd^4yd^{4}z_1d^4z'_1{d^{4}z_2d^4z'_2}\Delta^{d\sigma\rho;\sigma'\rho'}_\Phi(x,y;z_1,z'_1,z_2,z'_2)
\frac{1}{N_c^2}\bar{\psi}^{\sigma}(z_1)\tilde{\bar{\eta}}(x){\psi}^{\rho}(z'_1)\bar{\psi}^{\sigma'}(z_2)\tilde{\eta}(y){\psi}^{\rho'}(z'_2)\right\},
\end{eqnarray}
where $\tilde{\eta}_{\alpha\beta}=\eta^i\lambda^{i}_{\alpha\beta}$, $\tilde{\bar{\eta}}_{\alpha\beta}=\bar{\eta}^i\lambda^{i}_{\alpha\beta}$ and
\begin{eqnarray}
&&\Delta^a_\Phi(x,y)=\sum^{\infty}_{n=0}{\int}d^{4}x_1d^4x'_1\cdots{d^{4}x_nd^4x'_n}
\frac{(-1)^ni^nN_c^n}{n!} \mathcal{G}_{\rho_1\cdots\rho_n}^{(na)\sigma_1\cdots \sigma_n}(x,y;x_1,x'_1\cdots,x_n,x'_n)\Phi^{\sigma_1\rho_1}(x_1,x'_1)\cdots\Phi^{\sigma_n\rho_n}(x_n,x'_n),\nonumber\\
&&
\end{eqnarray}
\begin{eqnarray}
\Delta^{b\sigma\rho}_\Phi(x,y;z_1,z'_1)=&&\sum^{\infty}_{n=0}{\int}d^{4}x_1d^4x'_1\cdots{d^{4}x_nd^4x'_n}
\frac{(-1)^ni^{n+1}N_c^n}{(n+1)!} \mathcal{G}_{\rho\rho_1\cdots\rho_n}^{((n+1)b)\sigma\sigma_1\cdots \sigma_n}(x,y;z_1,z'_1,x_1,x'_1\cdots,x_n,x'_n)\nonumber\\
&&\times\Phi^{\sigma_1\rho_1}(x_1,x'_1)\cdots\Phi^{\sigma_n\rho_n}(x_n,x'_n),
\end{eqnarray}
\begin{eqnarray}
\Delta^{c\sigma\rho}_\Phi(x,y;z_1,z'_1)=&&\sum^{\infty}_{n=0}{\int}d^{4}x_1d^4x'_1\cdots{d^{4}x_nd^4x'_n}
\frac{(-1)^ni^{n+1}N_c^n}{(n+1)!} \mathcal{G}_{\rho\rho_1\cdots\rho_n}^{((n+1)c)\sigma\sigma_1\cdots \sigma_n}(x,y;z_1,z'_1,x_1,x'_1\cdots,x_n,x'_n)\nonumber\\
&&\times\Phi^{\sigma_1\rho_1}(x_1,x'_1)\cdots\Phi^{\sigma_n\rho_n}(x_n,x'_n),
\end{eqnarray}
and
\begin{eqnarray}
\Delta^{d\sigma\rho;\sigma'\rho'}_\Phi(x,y;z_1,z'_1,z_2,z'_2)=&&\sum^{\infty}_{n=0}{\int}d^{4}x_1d^4x'_1\cdots{d^{4}x_nd^4x'_n}
\frac{(-1)^ni^{n+2}N_c^{n+2}}{(n+2)!} \nonumber\\
&&\hspace*{-1cm}\times\mathcal{G}_{\rho\rho'\rho_1\cdots\rho_n}^{((n+2)c)\sigma\sigma'\sigma_1\cdots \sigma_n}(x,y;z_1,z'_1,z_2,z'_2,x_1,x'_1\cdots,x_n,x'_n)
\Phi^{\sigma_1\rho_1}(x_1,x'_1)\cdots\Phi^{\sigma_n\rho_n}(x_n,x'_n).
\end{eqnarray}

Now the Generating functional can be written as
\begin{eqnarray}
Z[J,{\cal I},\bar{I},I,\bar{\eta}^i,\eta^i]&=&\int{\cal D}\psi{\cal D}\bar{\psi}{\cal D}\Phi{\cal D}\Pi\exp i\bigg\{\int d^{4}x\{
\bar{\psi}(i\partial\!\!\!/+J-\Pi){\psi}+\bar{I}\psi+\bar{\psi}I\}+\int d^{4}xd^{4}x'N_c \Phi^{\sigma\rho}(x,x')\Pi^{\sigma\rho}(x,x')\nonumber\\
&&\hspace*{-1cm}+\sum^{\infty}_{n=2}{\int}d^{4}x_1\cdots{d^4}x_{n}
d^{4}x_{1}'\cdots{d^4}x_{n}'N_c\frac{(-i)^n (N_cg^2)^{n-1}}{n!}\bar{G}^{\sigma_1\cdots\sigma_n}_{\rho_1\cdots\rho_n}(x_1,x'_1,\cdots,x_n,x'_n)\nonumber\\
&&\times\Phi^{\sigma_1\rho_1}(x_1 ,x'_1)\cdots\Phi^{\sigma_n\rho_n}(x_n ,x'_n)\nonumber\\
&&\hspace*{-1cm}+\int d^4x_1d^4x_1'd^4x_2'\bar{\psi}^{\sigma}_{\alpha}(x_1')\tilde{\cal I}^{\mu}_{\alpha\beta}(x_1)\Delta^{\sigma\rho}_{\Phi,\mu}(x_1,x_1',x_2'){\psi}^{\rho}_{\beta}(x'_2)
+O({\cal I}^2)\bigg\}\nonumber\\
&&\hspace*{-1cm}\times\exp i\left\{{\int}d^4xd^4y\Delta^a_\Phi(x,y)\bar{\eta}^i(x)\eta^i(y)+{\int}d^4xd^4yd^{4}z_1d^4z'_1\Delta^{b\sigma\rho}_\Phi(x,y;z_1,z'_1)
\bar{\psi}^{\sigma}(z_1)\tilde{\bar{\eta}}(x)\tilde{\eta}(y){\psi}^{\rho}(z'_1)\right.\nonumber\\
&&\hspace*{-1cm}-{\int}d^4xd^4yd^{4}z_1d^4z'_1\Delta^{c\sigma\rho}_\Phi(x,y;z_1,z'_1)
\bar{\psi}^{\sigma}(z_1)\tilde{\eta}(y)\tilde{\bar{\eta}}(x){\psi}^{\rho}(z'_1)\nonumber\\
&&\hspace*{-1cm}\left.+{\int}d^4xd^4yd^{4}z_1d^4z'_1{d^{4}z_2d^4z'_2}\Delta^{d\sigma\rho;\sigma'\rho'}_\Phi(x,y;z_1,z'_1,z_2,z'_2)
\bar{\psi}^{\sigma}(z_1)\tilde{\bar{\eta}}(x){\psi}^{\rho}(z'_1)\bar{\psi}^{\sigma'}(z_2)\tilde{\eta}(y){\psi}^{\rho'}(z'_2)\right\}.
\end{eqnarray}
Introducing
\begin{eqnarray}
\int {\cal D}\Phi_{\bar{\eta}}{\cal D}\Phi_{\eta}~\delta \bigg(N_c\Phi_{\bar{\eta}}^{\sigma\rho}(x,x';y)-
\bar{\psi}^{\sigma}(x)\tilde{\bar{\eta}}(y){\psi}^{\rho}(x')\bigg)\delta \bigg(N_c\Phi_{\eta}^{\sigma\rho}(x,x';y)-
\bar{\psi}^{\sigma}(x)\tilde{\eta}(y){\psi}^{\rho}(x')\bigg),\label{PhiJ}
\end{eqnarray}
where
\begin{eqnarray}
\delta \bigg(N_c\Phi_{\bar{\eta}}^{\sigma\rho}(x,x';y)-\bar{\psi}^{\sigma}(x)\tilde{\bar{\eta}}(y){\psi}^{\rho}(x')\bigg)\sim\int{\cal D}\Pi_{\bar{\eta}} e^{i\int d^4xd^4x'd^4y
\big(N_c\Phi_{\bar{\eta}}^{\sigma\rho}(x,x',y)-\bar{\psi}^{\sigma}(x)\tilde{\bar{\eta}}(y){\psi}^{\rho}(x')\big)\cdot\Pi_{\bar{\eta}}^{\sigma\rho}(x,x';y)}
\nonumber
\end{eqnarray}
and following the similar procedure as that in section \uppercase\expandafter{\romannumeral2}, we obtain
\begin{eqnarray}
&&Z[J,{\cal I},\bar{I},I,\bar{\eta}^i,\eta^i]\nonumber\\
&&\hspace*{-1cm}=\int{\cal D}\psi{\cal D}\bar{\psi}{\cal D}\Phi{\cal D}\Pi{\cal D}\Phi_{\bar{\eta}}{\cal D}\Phi_{\eta}{\cal D}\Pi_{\bar{\eta}}{\cal D}\Pi_{\eta}\nonumber\\
&&\hspace*{-1cm}\times\exp i\bigg\{\int d^{4}xd^4y\big\{
\bar{\psi}[(i\partial\!\!\!/+J-\Pi)\delta(x-y)+\tilde{\cal I}\Delta_\Phi+\Delta^{b}_\Phi\tilde{\bar{\eta}}\tilde{\eta}-\Delta^{c}_\Phi\tilde{\eta}\tilde{\bar{\eta}}+\tilde{\bar{\eta}}\Pi_{\bar{\eta}}+\Pi_{\eta}\tilde{\eta}]{\psi}\nonumber\\
&&\hspace*{-1cm}+\int d^{4}x\{\bar{I}\psi+\bar{\psi}I\}+\int d^{4}xd^{4}x'N_c \Phi^{\sigma\rho}(x,x')\Pi^{\sigma\rho}(x,x')+{\int}d^4xd^4y\Delta^a_\Phi(x,y)\bar{\eta}^i(x)\eta^i(y)\nonumber\\
&&\hspace*{-1cm}+N_c\int d^4xd^4x'd^4y\left[\Phi_{\bar{\eta}}^{\sigma\rho}(x,x',y)\Pi_{\bar{\eta}}^{\sigma\rho}(x,x';y)+\Phi_{\eta}^{\sigma\rho}(x,x',y)\Pi_{\eta}^{\sigma\rho}(x,x';y)\right]\nonumber\\
&&\hspace*{-1cm}+\sum^{\infty}_{n=2}{\int}d^{4}x_1\cdots{d^4}x_{n}
d^{4}x_{1}'\cdots{d^4}x_{n}'N_c\frac{(-i)^n (N_cg^2)^{n-1}}{n!}\bar{G}^{\sigma_1\cdots\sigma_n}_{\rho_1\cdots\rho_n}(x_1,x'_1,\cdots,x_n,x'_n)
\Phi^{\sigma_1\rho_1}(x_1 ,x'_1)\cdots\Phi^{\sigma_n\rho_n}(x_n ,x'_n)\nonumber\\
&&\hspace*{-1cm}\left.+{\int}d^4xd^4yd^{4}z_1d^4z'_1{d^{4}z_2d^4z'_2}\Delta^{d\sigma\rho;\sigma'\rho'}_\Phi(x,y;z_1,z'_1,z_2,z'_2)
\Phi_{\bar{\eta}}^{\sigma\rho}(z_1,z'_1;y)\Phi_{\eta}^{\sigma'\rho'}(z_2,z'_2;x)\right\}.
\end{eqnarray}
Integrating out $\psi$ and $\bar{\psi}$, keeping up to $\eta\bar{\eta}$-order terms, we obtain
\begin{eqnarray}
Z'[J,{\cal I},\bar{I},I,\bar{\eta}^i,\eta^i]&\equiv&\lim_{N_c\rightarrow\infty}Z[J,{\cal I},\bar{I},I,\bar{\eta}^i,\eta^i]\nonumber\\
&=&\exp i\bigg\{-iN_c\mathrm{Tr}\ln\big[
(i\partial\!\!\!/+J-\Pi_c)\big]-i\mathrm{Tr}'\big[(i\partial\!\!\!/+J-\Pi_c)^{-1}\big(
\Delta^{b}_{\Phi_c}\tilde{\bar{\eta}}\tilde{\eta}-\Delta^{c}_{\Phi_c}\tilde{\eta}\tilde{\bar{\eta}}\big)\big]\nonumber\\
&&
+i\mathrm{Tr}'\big[(i\partial\!\!\!/+J-\Pi_c)^{-1}\tilde{\bar{\eta}}\Pi_{\bar{\eta} c}(i\partial\!\!\!/+J-\Pi_c)^{-1}\tilde{\eta}\Pi_{\eta c}\big]\nonumber\\
&&-\bar{I}\big[
(i\partial\!\!\!/+J-\Pi_c)^{-1}\big]I+\bar{I}\big[(i\partial\!\!\!/+J-\Pi_c)^{-1}\big(\tilde{\cal I}\Delta_{\Phi_c}+
\Delta^{b}_{\Phi_c}\tilde{\bar{\eta}}\tilde{\eta}-\Delta^{c}_{\Phi_c}\tilde{\eta}\tilde{\bar{\eta}}\big)(i\partial\!\!\!/+J-\Pi_c)^{-1}\big]I\nonumber\\
&&-\bar{I}\big[
(i\partial\!\!\!/+J-\Pi_c)^{-1}\tilde{\bar{\eta}}\Pi_{\bar{\eta} c}(i\partial\!\!\!/+J-\Pi_c)^{-1}\Pi_{\eta c}\tilde{\eta}(i\partial\!\!\!/+J-\Pi_c)^{-1}\big]I\nonumber\\
&&-\bar{I}\big[
(i\partial\!\!\!/+J-\Pi_c)^{-1}\Pi_{\eta c}\tilde{\eta}(i\partial\!\!\!/+J-\Pi_c)^{-1}\tilde{\bar{\eta}}\Pi_{\bar{\eta} c}(i\partial\!\!\!/+J-\Pi_c)^{-1}\big]I\nonumber\\
&&+{\int}d^4xd^4y\Delta^a_{\Phi_c}(x,y)\bar{\eta}^i(x)\eta^i(y)+N_c\int d^{4}xd^{4}x' \Phi_c^{\sigma\rho}(x,x')\Pi_c^{\sigma\rho}(x,x')\nonumber\\
&&+N_c\int d^4xd^4x'd^4y\left[\Phi_{\bar{\eta}c}^{\sigma\rho}(x,x',y)\Pi_{\bar{\eta} c}^{\sigma\rho}(x,x';y)+\Pi_{\eta c}^{\sigma\rho}(x,x';y)\Phi_{\eta c}^{\sigma\rho}(x,x',y)\right]\nonumber\\
&&+N_c\sum^{\infty}_{n=2}{\int}d^{4}x_1 d^{4}x_{1}'\cdots{d^4}x_{n}{d^4}x_{n}'\frac{(-i)^n (N_cg^2)^{n-1}}{n!}\nonumber\\
&&\times\bar{G}^{\sigma_1\cdots\sigma_n}_{\rho_1\cdots\rho_n}(x_1,x'_1,\cdots,x_n,x'_n)
\Phi_c^{\sigma_1\rho_1}(x_1 ,x'_1)\cdots\Phi_c^{\sigma_n\rho_n}(x_n ,x'_n)\nonumber\\
&&+{\int}d^4xd^4yd^{4}z_1d^4z'_1{d^{4}z_2d^4z'_2}\Delta^{d\sigma\rho;\sigma'\rho'}_{\Phi_c}(x,y;z_1,z'_1,z_2,z'_2)
\Phi_{\bar{\eta} c}^{\sigma\rho}(z_1,z'_1;y)\Phi_{\eta c}^{\sigma'\rho'}(z_2,z'_2;x)\bigg\},~~~\label{ZprimGhost}
\end{eqnarray}
where $\Phi_{\eta c}, \Phi_{\bar{\eta} c},\Pi_{\eta c}, \Pi_{\bar{\eta} c}$ are the expectation values of fields $\Phi_{\eta }, \Phi_{\bar{\eta} },\Pi_{\eta }, \Pi_{\bar{\eta} }$ respectively.

%%%%%%%%%%%%%%%%%%%%%%%%%%%%%%%%%%%%%%%%%%%%%%%%%%%%%%
\section*{Acknowledgments}
This work was supported by the National Science Foundation of China (NSFC) under Grant No. 11475092.

\end{document}